\documentclass[journal]{IEEEtran}
\usepackage{xcolor,soul,framed} 
\usepackage{cite}
\usepackage{amsmath,amssymb,amsfonts}
\usepackage{graphicx}
\usepackage{textcomp}
\usepackage{dcolumn}
\usepackage{bm}
\usepackage{subfigure} 
\usepackage{pict2e}
\usepackage{epstopdf}
\usepackage{arydshln}
\usepackage{fancyhdr}
\usepackage{booktabs}
\usepackage{tabularx}
\usepackage{comment}
\usepackage{mathtools}
\usepackage{tikz}
\usepackage{url}
\usepackage{bm}
\usepackage{tasks}
\usepackage{colortbl}
\usepackage[table]{xcolor}
\usepackage{array}
\usepackage{caption}
\usepackage{pifont}

\definecolor{yellow2}{RGB}{230, 230, 0}
\definecolor{orange}{RGB}{255, 180, 50}
\definecolor{blue}{RGB}{0, 155, 230}
\definecolor{pink}{RGB}{255, 182, 193}
\definecolor{green}{RGB}{0, 130, 80}
\definecolor{brown}{RGB}{210, 105, 30}

\begin{document}

\title{Gamification in Radiocommunications: A Board Game Approach to Boost Engagement and Learning}

\author{Ana S. Domenech, Antonio Alex-Amor 
\thanks{This work was supported in part by the BBVA Foundation's Leonardo Grant for Scientific Research and Cultural Creation 2025, and in part by the Grant PID2024-155167OA-I00 funded by MICIU/AEI/10.13039/501100011033/FEDER, UE. The BBVA Foundation is not responsible for the opinions, comments, and content included in the project and/or the results derived from it, which are the sole and absolute responsibility of their authors.  }
\thanks{Ana S. Domenech is with the Department of Information Technologies, Institute of Technology, Universidad San Pablo-CEU, CEU Universities, Campus Monteprincipe, Avenida Monteprincipe, Boadilla del Monte, 28660 Madrid, Spain (email: ana.sanmartindomenech@ceu.es)}
\thanks{A. Alex-Amor is with the Department of Electronic and Communication Technology, RFCAS Research Group,  Universidad Autónoma de Madrid, 28049 Madrid, Spain (email: antonio.alex@uam.es)}
}

\maketitle

\newcommand*{\bigs}[1]{\vcenter{\hbox{\scalebox{2}[8.2]{\ensuremath#1}}}}

\newcommand*{\bigstwo}[1]{\vcenter{\hbox{ \scalebox{1}[4.4]{\ensuremath#1}}}}

\begin{abstract}
    
Courses in electromagnetism and related technical subjects are often dominated by lecture-heavy instruction and complex mathematical concepts, which can make it difficult for students to stay engaged. This is particularly problematic in today’s hyper-digitalized society, where constant screen exposure and shortened attention spans challenge traditional learning methods. While computer-based tools and hands-on laboratories offer some pedagogical improvements, they often fall short in terms of interactivity, dynamism, adaptiveness, and student engagement. In an effort to enrich the learning experience and boost student motivation, we have created a gamified learning activity for the undergraduate course ``Radiocommunications"—commonly referred to as Antennas and Propagation in other institutions—, implemented in the form of a question-based board game. The activity, carried out over three academic years, is fully aligned with the course syllabus and encourages active learning, healthy competition, and collaborative problem-solving. Custom-made materials—including a game board, 270 question cards, wildcards, and incentive-based rewards—were developed specifically for this purpose. Qualitative results from a student survey, together with statistical evidence from hypothesis testing, suggest that the activity enhances conceptual understanding, helps students connect ideas across related subjects, and contributes to a more motivating and enjoyable learning experience.

\end{abstract}


\IEEEpeerreviewmaketitle

\section{Introduction}

Teaching electromagnetism is always a challenging task, from the fundamentals physics  learned in high school and early college courses to more advanced applied concepts seen in optics, antenna theory, and radiocommunication courses \cite{Sadiku1986, SuarezEM2024, WuRF2016,  PuertoOptics2022}. This is not so much due to understanding the main physical concepts in electromagnetism, but rather to the complexity of the mathematics involved \cite{PepperMath2012, BollenMath2015}. The situation becomes particularly problematic as the degree progresses, making it essential for the student to handle with ease advanced mathematical topics—such vector algebra, multivariable calculus, and differential equations—for fully understanding Maxwell's equations and their engineering applications. Frequently, this situation leads to the student becoming frustrated and unable to keep up with the class, which is reflected in the high number of students who typically fail these courses \cite{Thesis2011}. 

In fact, it is a common practice across many educational institutions that basic and more advanced electromagnetism courses are often delivered in a highly theoretical and traditional manner. While this approach has historical merit, it presents a significant challenge in today’s academic landscape. The current generation of students—particularly those belonging to Generation Z and the emerging Generation Alpha—have grown up immersed in a hyper-digitalized environment \cite{Alruthaya2021, Ameen2023, Swargiary2024}. Constant exposure to screens through smartphones, tablets, laptops, and other digital devices has profoundly influenced their cognitive development, particularly in terms of attention span, information processing, and memory retention \cite{Onyeaka2022, Panjeti2023}. As a result, these students often struggle to engage deeply with abstract and conceptually dense material, such as that encountered in (applied) electromagnetism courses. Their ability to maintain sustained concentration and perform in-depth, focused study is increasingly compromised, especially within rigid, lecture-heavy educational settings that do not accommodate their evolving cognitive and learning preferences \cite{Risko2013, Farley2013, Lodge2019, Jordan2019}.

It is precisely for the aforementioned reasons that innovative learning schemes, processes, and activities are needed to engage and motivate students in the challenging domain of (applied) electromagnetism. Following on from what was discussed earlier, the implementation of computer-based activities can significantly enhance students' ability to visualize and interpret electromagnetic (EM) fields, waves, and light-matter interactions in complex media  \cite{Cheng_Computer2003, Branislav_Computer2019, BaitSuwailam_Computer2023}. These activities can be complemented in turn with hands-on assignments and laboratories that directly put into practice the knowledge acquired by the student during the present and past courses \cite{Tornero2011, Ramon_Lab2012, Aliakbarian_Lab2014}. These activities improves conceptual understanding and development of practical skills, which is of great usefulness for the student with regard to finding a future job either in industry or academia.    

While computer-based activities and hands-on laboratories represent a significant step forward from purely theoretical instruction, it is also true that these approaches often lack the interactive elements—such as immediate feedback, adaptive progression, social interaction, and reward systems—that characterize more contemporary, student-centered strategies like serious games and gamified learning activities \cite{Sailer_Gamification2020, ZhaoGame2022, Szilagyi2025, Li_Gamification2023, Buenadicha2025}. These features are particularly effective in sustaining student motivation, encouraging active participation, and fostering a sense of achievement, all of which contribute to deeper and more enduring learning experiences. Such game-based approaches can serve as a valuable complement to conventional methods, offering students a more dynamic and motivating environment in which to explore complex electromagnetic concepts seen in radio, optics and photonics sciences \cite{Perez_Games2021, Richardson_Games2018, Vrignat_Games2025, Gaurina2025}. 

With the aim of engaging students  into the undergraduate 3$^\mathrm{rd}$-year course entitled ``Radiocommunications"—commonly referred to as \textit{Antenna Theory \& Propagation} in other institutions—taught at Universidad San Pablo-CEU (Madrid, Spain), we have created a non-profit gamified learning activity inspired from the popular Trivial Pursuit game. The activity is designed to reinforce the key concepts taught during the course. It directly aligns with the course syllabus, which covers five main instructional units: 1) review of electromagnetic fields and waves, 2) basic antenna parameters, 3) propagation, 4) wire antennas, and 5) apertures and arrays. These units serve as the foundation for the game’s question categories, ensuring complete integration between the instructional content and the learning tool. 

In the following, we will detail the context of the gamified activity and all the materials that were specifically created for it, including the game board, questions cards and wildcards, and special gifts. For reference and reuse, the reader  will have access in the supplementary material to the full list 270 custom-made question cards. Finally, we will share our experience and learnings across the three editions of the activity that have taken place, including the results from an anonymous student survey conducted at the conclusion of each edition, as well as a statistical analysis on the grades of the students who participated and did not participate in the game.

\section{Materials and Methods}

The activity has been carried out over \emph{three editions}, with a duration of two hours, held during the fall semesters of 2023, 2024 and 2025. It was strategically scheduled at the end of the semester, before final exams, aligned with the \textit{Ugly Sweater Day}—the third Friday of December. This date contributed to create a festive, informal and relaxed atmosphere which contrasted with the usual academic routine. 

In general terms, the activity operates as follows. The different teams that participate must collect  six colored pieces. Every space has a color associated with a specific category out of a total of six. Randomly, a question card of the same color/category is drawn. If the team answers the question correctly, they roll the die again.  If the team fails, the next team takes the turn. Out of all the spaces on the board, only six of them award pieces. All the teams try to reach these six special spaces. The team that collects all six pieces wins. 

\begin{figure}[!t]
	\centering
\subfigure{\includegraphics[width= 1\columnwidth]{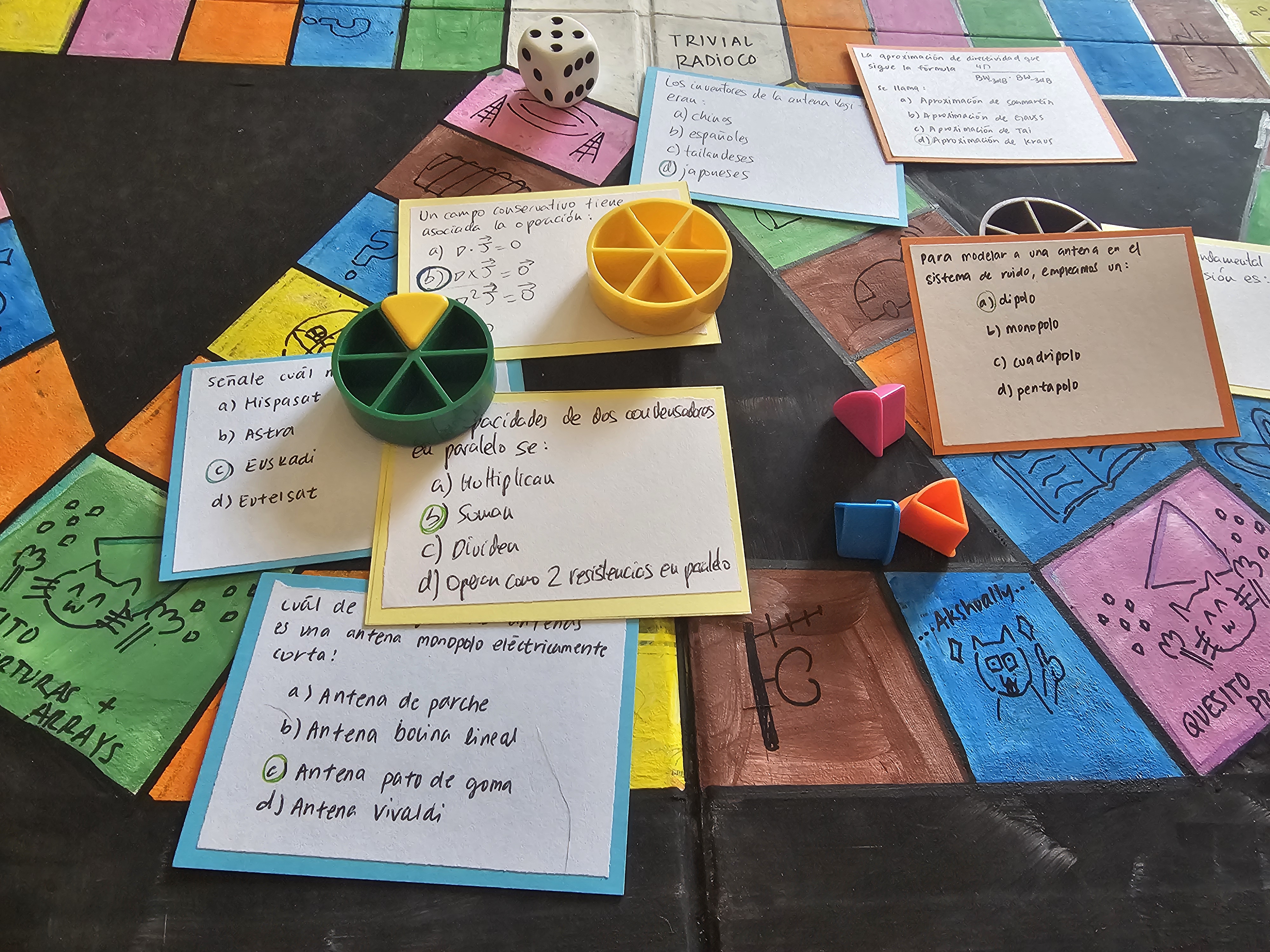}}
	\caption{Illustration of the hand-painted game board and some of the original handmade question cards in Spanish. }
	\label{fig1}
\end{figure}

Fig. \ref{fig1} illustrates the handmade game board and some of the  question cards created by the authors and specifically tailored for the activity. As can be seen, the handmade cards, which follow the color code described below, were originally created in Spanish, since Spanish is the language in which the course ``Radiocommunications" is taught at Universidad San Pablo-CEU. 

\subsection{Handmade Question Cards}

All questions were derived exclusively from course materials, ensuring academic rigor and alignment with the intended learning outcomes. In total, 270 handmade question cards—following the color code below—were created and distributed into six categories as follows:

\begin{itemize}
    \item \emph{47 for Review of EM Fields and Waves (yellow \protect\tikz[baseline=-0.5ex]{\protect\draw[yellow2, very thick] (0,0) -- (0.5,0);})}: It represents the first unit of the ``Radiocommunications" course. In this unit, the students review the fundamentals of electromagnetics seen in previous courses, such as electrostatics, Maxwell's equations, electric and magnetic fields, conduction and displacement currents, boundary conditions,  polarization, plane wave propagation, transmission lines, and waveguides. An example of handmade question card from this category is ``\emph{Two point charges $q_1$ and $q_2$ separated a distance $r$ suffer a force proportional to: a) $1/r$, \textbf{b)} $\mathbf{1/r^2}$, c) $1/r^3$, d) $1/r^4$".}
    
    \item \emph{48 for Basic Antenna Parameters (orange \protect\tikz[baseline=-0.5ex]{\protect\draw[orange, very thick] (0,0) -- (0.5,0);})}:
    This category covers the second unit of the course, where fundamental antenna parameters such as directivity, gain, radiation resistance, near-field and far-field regions, and antenna efficiency are defined. An example of handmade question card from this category is ``\emph{The power radiated per unit of solid angle is called: a) Directivity, b) Flux density , \textbf{c) Radiation intensity}, d) Radiation resistance."}
    
    \item \emph{48 for Propagation (pink \protect\tikz[baseline=-0.5ex]{\protect\draw[pink, very thick] (0,0) -- (0.5,0);})}:
    This category covers the third unit of the course, where the influence of the terrestrial environment on wave propagation is identified and appropriately modeled, such as differentiation and formulation of the propagation mechanisms of surface waves, ionospheric waves and space waves in radio links. An example of handmade question card from this category is ``\emph{In multi-path fading, if a dominant component is not present, the following statistical distribution is typically used: a) Rice, b) Binomial, c) Normal, \textbf{d) Rayleigh.}"}
    
    \item \emph{47 for Wire Antennas (brown \protect\tikz[baseline=-0.5ex]{\protect\draw[brown, very thick] (0,0) -- (0.5,0);})}: This category covers the fourth unit of the course, where the radiation characteristics of linear wire antennas such as the infinitesimal/short/half-wavelength dipoles, the quarter-wave monopole on a ground plane, and loops are analyzed. An example of handmade question card from this category is ``\emph{The antennas placed over a conductor plane are analyzed through the: a) Method of reflections, b) Method of reflectivity, c) Maxwell's method, \textbf{d) Method of images}."}
     
    \item \emph{28 for Aperture Antennas and Arrays (green \protect\tikz[baseline=-0.5ex]{\protect\draw[green, very thick] (0,0) -- (0.5,0);})}: It covers the fifth and last unit of the course, where the principle of equivalent currents is applied to the analysis of aperture and horn antennas. Moreover, reflector antennas and arrays are also discussed. An example of handmade question card from this category is ``\emph{Increasing the size of the aperture antenna also increases its: \textbf{a) Directivity}, b) Efficiency, c) Cutoff frequency, d) Our patience."}
    
    \item \emph{52 for Science Curiosities (blue \protect\tikz[baseline=-0.5ex]{\protect\draw[blue, very thick] (0,0) -- (0.5,0);})}: We consider of interest to add a sixth category named ``Science Curiosities" to the five previous categories based on the five instructional units of the course. We believe that it is always worthwhile to go beyond theoretical content and learn about the context of relevant scientists in the field and exploring the technologies developed in connection with the subject. Thus, the category ``Science Curiosities" incorporates anecdotes, historical facts, and informal comments shared by professors during lectures. An example of handmade question card from this category is ``\emph{The inventors of the Yagi-Uda antenna were: a) Chinese, b) Spanish, c) Thai, \textbf{d) Japanese}."} 
\end{itemize}
The reader is referred to the Supplementary Material to see the full list of 270 questions used in the activity, translated from Spanish to English. 

In addition to the question cards,  \emph{wildcards} were introduced, thus creating a new strategic component to the game and sustaining a high level of engagement. A total of 12 handmade wildcards were designed, separated into two categories: special wildcards (multiple occurrences) and simple wildcards (only one occurrence):
\begin{itemize}
    \item (2x) \emph{Question-Stealing Wildcard}: Allows stealing a question from an opposing team before they answer.
    \item (2x) \emph{Question-Passing Wildcard}: Allows receiving a new question by ignoring the current one. Only applicable during the same turn and if the current question has not been answered.
    \item (2x) \emph{Phone-a-Friend Wildcard}: Allows calling another person to clarify doubts. The person can be outside the classroom or present.
    \item (1x) \emph{Error-Nullifying Wildcard}: Allows re-answering the same question after an incorrect attempt without losing the turn.
    \item (1x) \emph{Review Notes Wildcard}: Allows for consulting course materials for 3 minutes.
    \item (1x) \emph{Ask the audience Wildcard}: General vote of all players to clarify doubts.
    \item (1x) \emph{Steal Opponent Wildcard}: Allows swapping a player from your team with a player from an opposing team for an entire turn.
    \item (1x) \emph{Pass-to-Next team Wildcard}: Allows to pick a team to be skipped for one turn.
    \item (1x) \emph{Eliminate Two Options Wildcard}: Allows to remove 2 out of 4 options.
\end{itemize}

As amusing anecdotes related to the use of wildcards, two particularly memorable moments stand out. In the first, a student from the inaugural edition of the game used the "Phone-a-Friend" wildcard to call his brother—a fourth-year student who had passed the course the previous year. At the time, his brother was on an Erasmus exchange in Italy. He answered the call but, unfortunately, failed to answer the question correctly. In the second anecdote, one of the students teams initially got a question wrong and decided to use the "Error-Nullifying" wildcard, which allowed them to attempt the question again. However, just before they could re-answer, the professors team activated the "Question-Stealing" wildcard, taking over the question and answering it themselves. These anecdotes highlight how wildcards bring a fun and strategic twist to the game.

\subsection{Teams and Participants}

\begin{figure}[!t]
	\centering
\subfigure{\includegraphics[width= 1\columnwidth]{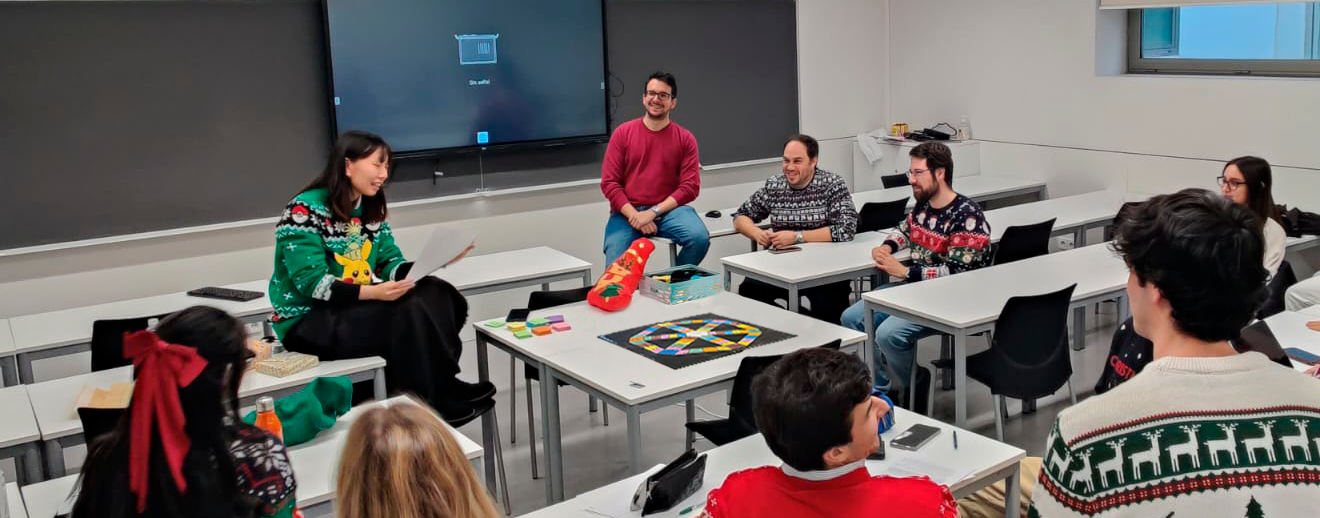}}
	\caption{A picture of some of the participants—students and professors—of the second edition (fall semester, 2024) playing the activity. The main instructor of the ``Radiocommunications" course appears on stage asking a question while the team members discuss what the correct answer is.} 
	\label{fig2}
\end{figure}

In the three editions of the activity, teams were organized into groups of four students. Teams of this size ensure that all members participate actively and, at the same time, there are not too many teams, which would slow down the game flow. The main professor of the course ``Radiocommunications", Ana S. Domenech, always acted as moderator/presenter during the activity, not being part of any team. 

To the student teams, we thought interesting to add an \emph{extra team} of three professors, professors that were closely related to the topics seen in the course. This format significantly increases healthy competitiveness, as students are highly motivated to outperform their instructors. In the first edition, we counted on the professors Antonio Alex-Amor (he taught the 2$^\mathrm{nd}$-year course ``Field and Waves"), Pablo Pérez-Tirador (he currently teaches the 4$^\mathrm{th}$-year course ``Radiocommunications Systems"), and Pedro de la Torre Luque (he taught the 1$^\mathrm{st}$-year course ``Waves, Electrostatics \& Thermodynamics"). In the second edition, we counted again on the professors Antonio Alex-Amor and Pablo Pérez-Tirador, and professor Rodrigo Rodríguez-Merino (he currently teaches the 2$^\mathrm{nd}$-year course ``Field and Waves"). In the third edition, we counted again on the professors Pablo Pérez-Tirador and Rodrigo Rodríguez-Merino, plus Javier Olivares Herrador (he currently teaches the 2$^\mathrm{nd}$-year course ``Electromagnetism and Optics") who joined the activity. 

Fig. \ref{fig2} shows a picture of some of the participants (students and professors) of the second edition playing the game, gathered in the usual room where the Radiocommunications course is taught.

\begin{figure}[!t]
	\centering
\subfigure{\includegraphics[width= 0.9\columnwidth]{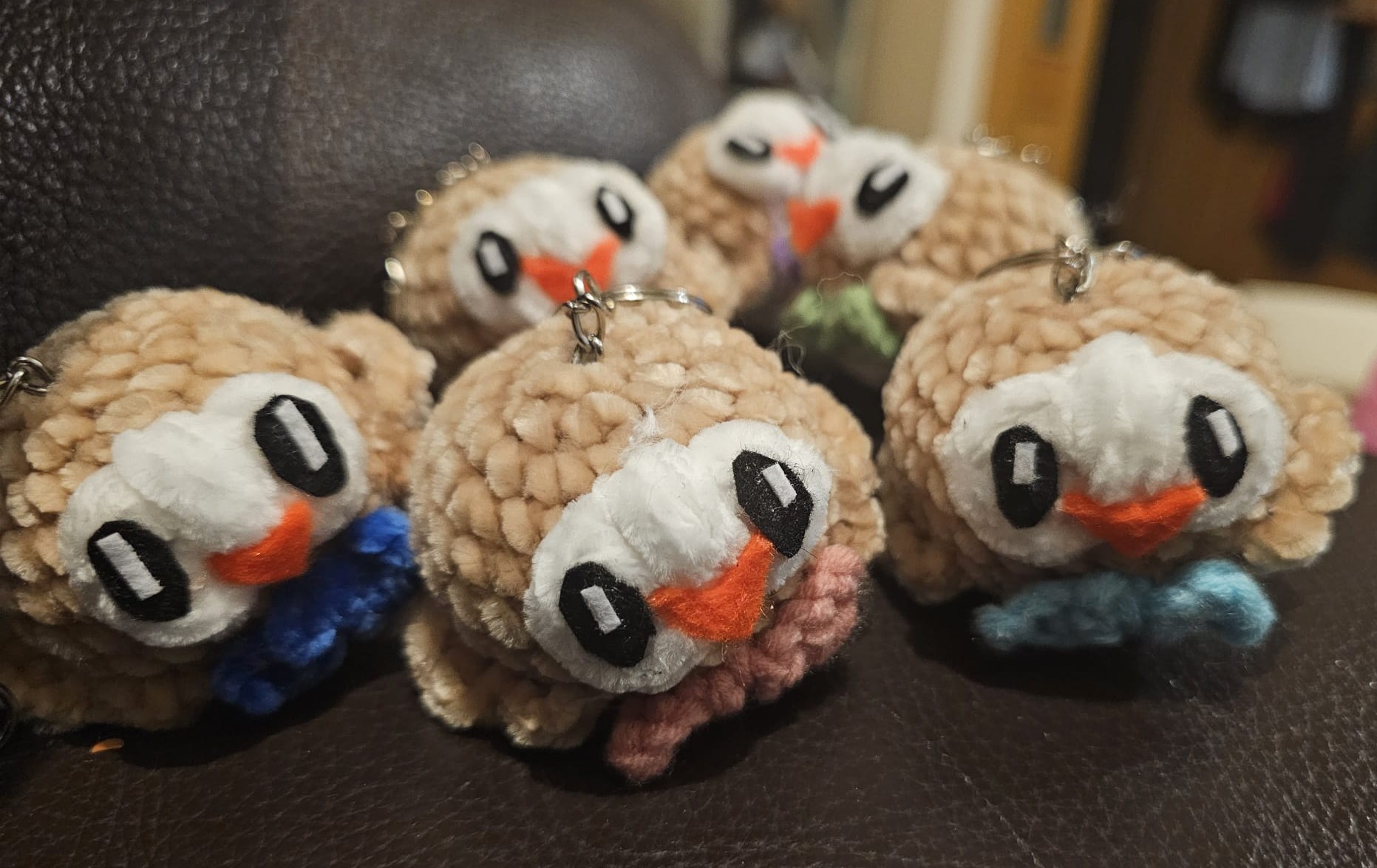}}
	\caption{Small handmade gifts given to the winning team.}
	\label{fig3}
\end{figure}

\subsection{Rewards}

Participation in the gamified activity is entirely optional, allowing students to choose whether or not to engage based on their individual preferences and schedules. Given that the activity takes place in December—coinciding with the lead-up to the final examination period—some students opted to prioritize studying over participation. Recognizing this, the course instructor makes a point of clearly communicating the indirect academic benefits associated with taking part in the activity. These include the opportunity to review and reinforce key course concepts, to organize and consolidate ideas in preparation for the final exam, and to engage in a more relaxed and informal learning environment during what is typically a high-stress time for students.

To further encourage participation, the activity incorporates a system of academic incentives. Specifically, extra credit is awarded to the winning team, providing a tangible and motivating benefit. In the context of the Spanish grading system, this translates into an additional point on the final  grade for the laboratory classes, which is scored on a scale from 0 to 10. This added incentive not only rewards high-performing participants but also underscores the academic relevance of the activity.

In addition to the academic rewards, the activity includes small gestures aimed at enhancing student morale and fostering a sense of community. All participants were offered sweets during the game. Moreover, the winning team received small handmade gifts by the main instructor of the course (see Fig. \ref{fig3}). These elements contribute to a positive and inclusive atmosphere, reinforcing the idea that the activity is not only academically valuable but also enjoyable and socially rewarding.

\begin{table*}[!t]
\centering
\caption{Final questionnaire where the students evaluate the gamified activity. The scale ranges from `1' = lowest to \linebreak `5' = highest. The numbers inside the boxes indicate the number of students that selected an option.}
\rowcolors{2}{gray!10}{white}
\renewcommand{\arraystretch}{1.4}
\begin{tabular}{
    >{\raggedright\arraybackslash}p{6cm}
    >{\centering\arraybackslash}p{1cm}
    >{\centering\arraybackslash}p{1cm}
    >{\centering\arraybackslash}p{1cm}
    >{\centering\arraybackslash}p{1cm}
    >{\centering\arraybackslash}p{1cm}
    >{\centering\arraybackslash}p{1cm}
}
\toprule
\rowcolor{gray!30}
\textbf{Grading} & \textbf{`1'} & \textbf{`2'} & \textbf{`3'} & \textbf{`4'} & \textbf{`5'} & \textbf{Average} \\
\midrule

1. The activity is appropriate and makes me enjoy the related course(s) more.  & $\boxed{0}$ & $\boxed{0}$ & $\boxed{0}$ & $\boxed{1}$ & $\boxed{23}$ & 4.96  \\

2. This activity has improved my understanding of the course.  & $\boxed{0}$ & $\boxed{0}$ & $\boxed{4}$ & $\boxed{7}$ & $\boxed{13}$ & 4.38 \\

3. This activity has incorporated knowledge from the course.             & $\boxed{0}$ & $\boxed{0}$ & $\boxed{0}$ & $\boxed{4}$ & $\boxed{20}$ & 4.83 \\

4. This activity has connected the course’s content with concepts from previous courses.        & $\boxed{0}$ & $\boxed{0}$ & $\boxed{3}$ & $\boxed{4}$ & $\boxed{17}$ & 4.58\\

5. This activity has improved my understanding of previous courses.        & $\boxed{0}$ & $\boxed{0}$ & $\boxed{6}$ & $\boxed{7}$ & $\boxed{11}$ & 4.21\\

6. This activity has motivated me to study/work in related areas.        & $\boxed{0}$ & $\boxed{0}$ & $\boxed{2}$ & $\boxed{8}$ & $\boxed{14}$ & 4.50\\

7. The proposal has allowed me to see a global perspective of the related subjects.        & $\boxed{0}$ & $\boxed{0}$ & $\boxed{3}$ & $\boxed{7}$ & $\boxed{14}$ & 4.56\\

8. The content of the activity has matched my level and training needs.        & $\boxed{0}$ & $\boxed{0}$ & $\boxed{1}$ & $\boxed{6}$ & $\boxed{17}$ & 4.67\\

9. Overall, the activity has made me enjoy the related courses more.        & $\boxed{0}$ & $\boxed{0}$ & $\boxed{0}$ & $\boxed{2}$ & $\boxed{22}$ & 4.92\\

10. The time employed for the activity has been sufficient  & $\boxed{1}$ & $\boxed{1}$ & $\boxed{1}$ & $\boxed{7}$ & $\boxed{14}$ & 4.33 \\

11. Please evaluate the activity.
& $\boxed{0}$ & $\boxed{0}$ & $\boxed{0}$ & $\boxed{3}$ & $\boxed{21}$ & 4.88\\

\bottomrule
\label{table1}
\end{tabular}
\end{table*}

\section{Evaluation of Results}

The evaluation of students' attitudes towards the activity has demonstrated a positive impact on the learning experience. In the three editions, students have consistently described the activity as pleasant and dynamic, showing active participation throughout the session. Beyond the recreational dimension, the activity is perceived as a tool for strengthening knowledge, helping students review and consolidate concepts that later appeared in the final exam. Furthermore, the playful format fostered healthy competitiveness, which, acted as a motivating factor that improved engagement and collaboration.

\subsection{Student Survey}

In addition to these observations, an anonymous survey composed of thirteen questions was distributed to objectively assess student perceptions. Of these, eleven were Likert-scale items (ranging from `1' = lowest to `5' = highest) designed to capture different aspects of the experience. As shown in Table \ref{table1}, the student's evaluations of the gamified activity are positive. All items scored above 4.21 (5.00 is the maximum), with particularly high agreement on the activity's ability to increase enjoyment of the related courses (items 1 and 9, average = 4.96 and 4.92, respectively) and its overall quality (item 11, average = 4.88). Items concerning integration with prior courses (item 5) and adequacy of time (item 10) received comparatively lower, though still strong, evaluations. These results suggest that the activity was both engaging and pedagogically valuable. They also indicate that, for a possible fourth edition in the future, it may be worth considering an extension of the activity's duration.

Questions 12 and 13, not shown in Table I, are open-ended questions: Q12. \textit{``What did you like the most about the activity and how you would improve it?"}; Q13. \textit{``Do you have any other suggestion or comment?"}. The qualitative responses revealed several recurring themes. Students highlighted the enjoyable and engaging nature of the activity, often highlighting the motivational role of competition in fostering deeper understanding of the subject matter (e.g., \textit{``It was an enjoyable activity that we all appreciated as a class, where the element of competition motivated us to make an effort to really understand the content"}). Some participants suggested allocating more time to the activity, emphasizing that the dedicated time was insufficient (\textit{``I would only improve it by dedicating a little more time."}). Additional student comments pointed to the pedagogical value of the activity as a comprehensive review tool (\textit{``It is a fun, engaging activity that helps review the entire subject, with questions that go beyond what would typically be asked in an exam"}), as well as appreciation for its innovative and enjoyable format (\textit{``It is very didactic and lively since we can even compete against our own professors"; ``Learning in a different way"}). Finally, students recommended incorporating more activities of this type and even making it a permanent component of the course (\textit{``There should be more activities like this one"; ``I would make it a permanent activity in the course; it was really cool"}).

\subsection{Statistical Analysis}

Although the qualitative benefits of the activity on students' engagement and development throughout the course have been observed, a relevant question arises at this point: does the gamified activity have a measurable positive impact on students' grades? Specifically, are the average grades of the students who participated in the activity, $\mu_p$, higher than those of the students who did not participate, $\mu_{dnp}$? While grades do not fully capture all aspects of the learning process—such as engagement, motivation, or the development of transversal skills—they remain an important \emph{measurable} indicator of students’ academic performance.

In the following, we test the hypothesis that students who participated in the activity have higher grades than those who did not, i.e., $\mu_p > \mu_{dnp}$. The two groups are independent, as the grades of students who participated do not influence the grades of students who did not. The sample sizes are different too. Moreover, the population variances of the two groups may differ, making Welch's t-test for two independent samples particularly appropriate in this scenario. 

\begin{figure}[!t]
	\centering
\subfigure{\includegraphics[width= 0.9\columnwidth]{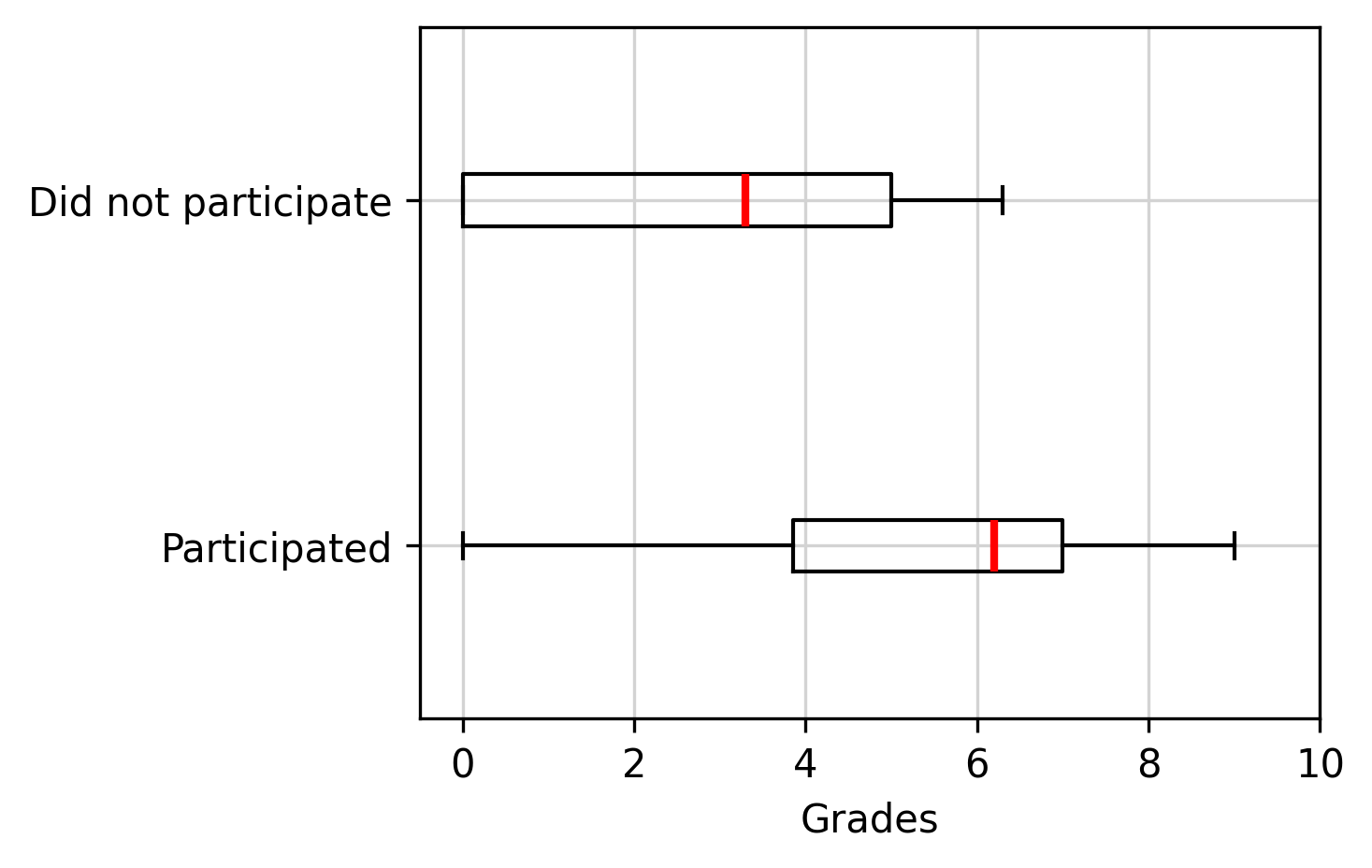}}
	\caption{Boxplot of grades for both groups: students who participated in the gamified activity and students who did not participate. The box represents the interquartile range (lower Q1 and upper Q3), the horizontal red line indicates the median, and the whiskers represent the range of the data.}
	\label{fig4}
\end{figure}

We analyzed a record of 60 students over three academic years who were registered for the activity. Of these, $n_p = 47$ students participated, while $n_{dnp} = 13$ did not. The average grade and standard deviation for participants were $\mu_p = 5.35$ and $s_p = 2.40$, respectively. For non-participants, the average and standard deviation were $\mu_{dnp} = 2.85$ and $s_{dnp} = 2.36$. Grades are evaluated on a scale from 0 (minimum) to 10 (honors), with a passing grade of 5. Students who do not take the exam automatically received a zero.  The boxplot in Fig.~\ref{fig4} provides a visual summary of the grades.

Using the data described above, Welch's t-test for two independent samples 
($H_0: \mu_p = \mu_{dnp}$; $H_A: \mu_p > \mu_{dnp}$) with a significance level of $\alpha = 0.05$ yielded a $p$-value of $p = 0.001566$. The corresponding test statistic, degrees of freedom, critical value, and Cohen's factor were $t_0 = 3.3708$, $\nu = 19.44$, $t_{\text{lim}} = 1.7271$, and $d = 1.0457$, respectively. Since $p = 0.001566 < \alpha$ (and $d > 0.8$), there is strong (and practically meaningful) statistical evidence that students who participated in the activity achieved higher grades than those who did not.

For reference, similar conclusions are obtained when students who did not take the exam are excluded from the analysis. In this case, the sample sizes are reduced to $n_p' = 43$ and $n_{dnp}' = 9$, and the prime notation is used to indicate that this is a distinct statistic. The mean and standard deviation for participants are $\mu_p' = 5.85$ and $s_p' = 1.82$, while for non-participants they are $\mu_{dnp}' = 4.12$ and $s_{dnp}' = 1.57$. The resulting $p$-value is $p' = 0.005933 < \alpha$, and the Cohen's $d$ factor is $d' = 0.9707 > 0.8$, indicating a large effect size. 

There is strong and practically meaningful statistical evidence that students who participated in the proposed gamified activity achieved higher grades than those who did not. While these results demonstrate a significant difference, it is important to emphasize that this does not necessarily imply a direct causal effect of the activity. It is possible that higher-performing students are more likely to engage in such activities, which further stimulate their knowledge and motivation. Conversely, the activity itself may have contributed to improved grades, or both factors may have acted together. In any case, the findings provide compelling evidence of a positive association between participation in the activity and academic performance, though causality cannot be conclusively established.

\section{Conclusion}

In this paper, we present our experience with a gamified learning activity inspired by the popular board game Trivial Pursuit, designed to engage and motivate students in the 3$^\mathrm{rd}$-year course ``Radiocommunications". The primary objective of this initiative is to foster a relaxed, student-friendly environment that encourages the review of key concepts from Radiocommunications and related electromagnetism courses such as Antenna Theory. The activity has been implemented during three consecutive fall semesters (2023, 2024 and 2025) and is closely aligned with the course syllabus, covering topics such as: 1) electromagnetic fields and waves, 2) fundamental antenna parameters, 3) radio wave propagation, 4) wire antennas, and 5) aperture antennas and arrays.

All materials used in the game—including the board, question cards, wildcards, and small incentive gifts—are handmade and specifically tailored for the activity. A total of 270 custom-designed questions are available as Supplementary Material for reference and reuse.

Student feedback across the three editions has consistently highlighted a positive learning experience, with participants describing the activity as enjoyable and dynamic. Responses to an anonymous voluntary survey indicated improvements in conceptual understanding, the ability to connect ideas across related subjects, and overall motivation. In addition, statistical analysis on the acquired data provides strong evidence that students who participated in the activity obtained higher final grades in the Radiocommunications course than those who did not. These results underscore the potential of game-based strategies to enhance engineering education, particularly in a digital era where interactive and learner-centered approaches are increasingly important.





\ifCLASSOPTIONcaptionsoff
  \newpage
\fi

\bibliographystyle{IEEEtran}
\bibliography{IEEEabrv, ref}



\pagebreak

\setcounter{equation}{0}
\setcounter{figure}{0}
\setcounter{table}{0}
\setcounter{page}{1}
\makeatletter
\renewcommand{\theequation}{S\arabic{equation}}
\renewcommand{\thefigure}{Supp.\arabic{figure}}
\renewcommand{\labelenumi}{\arabic{enumi}.}


\onecolumn
\begin{center}
\textrm{\huge Supplementary Material}
\end{center}

\vspace{0.2cm}

\begin{center}
\textrm{\LARGE Gamification in Radiocommunications: \\ A Board Game Approach to Boost Engagement and Learning}
\end{center}

\begin{center}
\textrm{\Large Ana S. Domenech$^{*}$, Antonio Alex-Amor$^\ddagger$}
\end{center}

\begin{center}
\textit{$*$Department of Information Technologies, Institute of Technology, Universidad San Pablo-CEU, CEU Universities, Campus Monteprincipe, Avenida Monteprincipe, Boadilla del Monte, 28660 Madrid, Spain.}
\end{center}

\begin{center}
\textit{$\ddagger$Department of Electronic and Communication Technology, RFCAS Research Group,  Universidad Autónoma de Madrid, 28049 Madrid, Spain.}
\end{center}


\section*{List of Handmade Questions Cards created for the Gamified Activity}

The 270 questions created for the activity and written on handmade cards are listed below. The correct answer is highlighted with \textbf{bold} letters.

The questions are divided into six categories. Each category is aligned with the five instructional units (plus the additional category ``Science Curiosities") of the 3$^\mathrm{rd}$-year course ``Radiocommunications" taught at Universidad  San Pablo-CEU (Madrid, Spain). To easily identify the six categories, we consider  the same color code used in the main text:

\begin{itemize}

    \item Wire Antennas (brown \protect\tikz[baseline=-0.5ex]{\protect\draw[brown, very thick] (0,0) -- (0.5,0);})

    \item Basic Antenna Parameters (orange \protect\tikz[baseline=-0.5ex]{\protect\draw[orange, very thick] (0,0) -- (0.5,0);})

    \item Aperture Antennas and Arrays (green \protect\tikz[baseline=-0.5ex]{\protect\draw[green, very thick] (0,0) -- (0.5,0);})

    \item Propagation (pink \protect\tikz[baseline=-0.5ex]{\protect\draw[pink, very thick] (0,0) -- (0.5,0);})
        
    \item Review of EM Fields and Waves (yellow \protect\tikz[baseline=-0.5ex]{\protect\draw[yellow2, very thick] (0,0) -- (0.5,0);})
    
    \item Science Curiosities (blue \protect\tikz[baseline=-0.5ex]{\protect\draw[blue, very thick] (0,0) -- (0.5,0);})
    
\end{itemize}

\begin{enumerate}
	 \item \emph{(\protect\tikz[baseline=-0.5ex]{\protect\draw[brown, very thick] (0,0) -- (0.5,0);})} The directivity of a 1.5$\lambda$ dipole antenna is (in linear units): 
	 \begin{tasks}(4)
		 \task 1.5
		 \task 1.64
		 \task 2.41
		 \task \textbf{2.17}
		 \end{tasks} 
	 \item \emph{(\protect\tikz[baseline=-0.5ex]{\protect\draw[brown, very thick] (0,0) -- (0.5,0);})} For an elemental dipole over a conductive plane, the radiation resistance is \_\_\_\_ compared to free space (for the same current):
	 \begin{tasks}(4)
		 \task The same
		 \task \textbf{ Double}
		 \task Triple
		 \task Quadruple
		 \end{tasks} 
	 \item \emph{(\protect\tikz[baseline=-0.5ex]{\protect\draw[brown, very thick] (0,0) -- (0.5,0);})} For an isotropic antenna over a conductive plane (c.p.) radiating Pt, what is the relationship of its Power flux density ($\phi$) compared to its free-space (f.s.) equivalent?: (c.p. = conductive plane; f.s. = free space; Pt = transmitted (radiated) power; $\phi$ = Power flux density) 
	 \begin{tasks}(4)
		 \task \bm{$\phi(c.p.) = 2\cdot\phi(f.s.)$}
		 \task $2\cdot\phi(c.p..) = \phi(f.s.)$
		 \task $\phi(c.p..) = \sqrt{2}\cdot\phi(f.s.)$
		 \task $\sqrt{2}\cdot\phi(c.p..) = \phi(f.s.)$
		 \end{tasks} 
	 \item \emph{(\protect\tikz[baseline=-0.5ex]{\protect\draw[brown, very thick] (0,0) -- (0.5,0);})} The directivity relative to an isotropic antenna is measured in:
	 \begin{tasks}(4)
		 \task \textbf{dBi}
		 \task dBa
		 \task dB
		 \task dBd
		 \end{tasks} 
	 \item \emph{(\protect\tikz[baseline=-0.5ex]{\protect\draw[brown, very thick] (0,0) -- (0.5,0);})} For an elemental dipole over a conductive plane, the total field is \_\_\_\_ compared to free space (for the same radiated power):
	 \begin{tasks}(4)
		 \task Same
		 \task Double
		 \task \bm{$\sqrt{2}$}
		 \task Quadruple
		 \end{tasks} 
	 \item \emph{(\protect\tikz[baseline=-0.5ex]{\protect\draw[brown, very thick] (0,0) -- (0.5,0);})} For an elemental monopole over a conductive plane, the radiation resistance is \_\_\_\_ compared to the equivalent dipole in free space (for the same current):
	 \begin{tasks}(4)
		 \task \textbf{Half}
		 \task The same
		 \task Double
		 \task $\sqrt{2}$
		 \end{tasks} 
	 \item \emph{(\protect\tikz[baseline=-0.5ex]{\protect\draw[brown, very thick] (0,0) -- (0.5,0);})} For an elemental dipole over a conductive plane, the total directivity is \_\_\_\_ compared to free space (for the same radiated power):
	 \begin{tasks}(4)
		 \task \textbf{The same}
		 \task Double
		 \task Triple
		 \task Quadruple
		 \end{tasks} 
	 \item \emph{(\protect\tikz[baseline=-0.5ex]{\protect\draw[brown, very thick] (0,0) -- (0.5,0);})} A false example of a balun is:
	 \begin{tasks}(4)
		 \task \textbf{Floating balun}
		 \task Bazooka balun
		 \task Ferrite balun
		 \task Split balun
		 \end{tasks} 
	 \item \emph{(\protect\tikz[baseline=-0.5ex]{\protect\draw[brown, very thick] (0,0) -- (0.5,0);})} For an elemental monopole over a conductive plane, the field is \_\_\_\_ compared to the equivalent dipole in free space (for the same current):
	 \begin{tasks}(4)
		 \task Half
		 \task The same
		 \task Double
		 \task \bm{$\sqrt{2}$}
		 \end{tasks} 
	 \item \emph{(\protect\tikz[baseline=-0.5ex]{\protect\draw[brown, very thick] (0,0) -- (0.5,0);})} The directive gain of a dipole of length $\lambda$ is: 
	 \begin{tasks}(4)
		 \task 1.5 (linear units)
		 \task 1.64 (linear units)
		 \task \textbf{2.41 (linear units)}
		 \task 2.17 (linear units)
		 \end{tasks} 
	 \item \emph{(\protect\tikz[baseline=-0.5ex]{\protect\draw[brown, very thick] (0,0) -- (0.5,0);})} The directivity of a short dipole is:
	 \begin{tasks}(4)
		 \task 1.64 dBi
		 \task 2.15 dBi
		 \task 1.5 dBi
		 \task \textbf{1.76 dBi}
		 \end{tasks} 
	 \item \emph{(\protect\tikz[baseline=-0.5ex]{\protect\draw[brown, very thick] (0,0) -- (0.5,0);})} The ARP uses as a reference antenna a:
	 \begin{tasks}(4)
		 \task Isotropic antenna
		 \task \textbf{Half-wave dipole}
		 \task Short vertical antenna
		 \task Quarter-wave monopole
		 \end{tasks} 
	 \item \emph{(\protect\tikz[baseline=-0.5ex]{\protect\draw[brown, very thick] (0,0) -- (0.5,0);})} The dipole whose length is much smaller than the wavelength is:
	 \begin{tasks}(4)
		 \task Perfect dipole
		 \task Quite short dipole
		 \task Small radiating dipole
		 \task \textbf{Short dipole}
		 \end{tasks} 
	 \item \emph{(\protect\tikz[baseline=-0.5ex]{\protect\draw[brown, very thick] (0,0) -- (0.5,0);})} For an elemental monopole over a conductive plane, the total radiated power is \_\_\_\_ compared to the equivalent dipole in free space (for the same current):
	 \begin{tasks}(4)
		 \task \textbf{Half}
		 \task The same
		 \task Double
		 \task $\sqrt{2}$
		 \end{tasks} 
	 \item \emph{(\protect\tikz[baseline=-0.5ex]{\protect\draw[brown, very thick] (0,0) -- (0.5,0);})} The antenna constructed by winding a conductor around a dielectric cylinder with a certain pitch angle is called:
	 \begin{tasks}(4)
		 \task Biconical antenna
		 \task Vivaldi antenna
		 \task Short vertical antenna
		 \task \textbf{Helical antenna}
		 \end{tasks} 
	 \item \emph{(\protect\tikz[baseline=-0.5ex]{\protect\draw[brown, very thick] (0,0) -- (0.5,0);})} The mutual impedance matrix model of a network with $N$ ports is used to measure:
	 \begin{tasks}(4)
		 \task The dipole’s vector potential $\vec{A}$
		 \task Dipole current
		 \task \textbf{Mutual coupling between the dipole and its image}
		 \task The field induced by the image field in the dipole
		 \end{tasks} 
	 \item \emph{(\protect\tikz[baseline=-0.5ex]{\protect\draw[brown, very thick] (0,0) -- (0.5,0);})} The current distribution pattern of a real dipole is approximately:
	 \begin{tasks}(4)
		 \task A quadrilateral
		 \task \textbf{A triangle}
		 \task A line
		 \task A circle
		 \end{tasks} 
	 \item \emph{(\protect\tikz[baseline=-0.5ex]{\protect\draw[brown, very thick] (0,0) -- (0.5,0);})} The current distribution pattern of an elemental dipole is a: 
	 \begin{tasks}(4)
		 \task \textbf{Constant}
		 \task Triangle
		 \task Sphere
		 \task Lie, it is a Hertz dipole, not elemental
		 \end{tasks} 
	 \item \emph{(\protect\tikz[baseline=-0.5ex]{\protect\draw[brown, very thick] (0,0) -- (0.5,0);})} The elemental dipole is also known by the name:
	 \begin{tasks}(4)
		 \task Watson dipole
		 \task Sherlock dipole
		 \task \textbf{Hertz dipole}
		 \task Maxwell dipole
		 \end{tasks} 
	 \item \emph{(\protect\tikz[baseline=-0.5ex]{\protect\draw[brown, very thick] (0,0) -- (0.5,0);})} In Yagi-Uda antennas, among the parallel dipoles, the one that is fed is: 
	 \begin{tasks}(4)
		 \task \textbf{One of them}
		 \task Two of them
		 \task None
		 \task All
		 \end{tasks} 
	 \item \emph{(\protect\tikz[baseline=-0.5ex]{\protect\draw[brown, very thick] (0,0) -- (0.5,0);})} The input resistance ($R_\text{in}$) of a thin, center-fed dipole of length $\lambda$ is: 
	 \begin{tasks}(4)
		 \task $36.6 \Omega$
		 \task $73.2 \Omega$
		 \task \bm{$\infty$}
		 \task $99.5 \Omega$
		 \end{tasks} 
	 \item \emph{(\protect\tikz[baseline=-0.5ex]{\protect\draw[brown, very thick] (0,0) -- (0.5,0);})} For dipoles parallel to a conductive plane and separated by $\lambda$/4, the power density is \_\_\_\_ compared to free space (for the same current):
	 \begin{tasks}(4)
		 \task Same
		 \task Double
		 \task Triple
		 \task \textbf{Quadruple}
		 \end{tasks} 
	 \item \emph{(\protect\tikz[baseline=-0.5ex]{\protect\draw[brown, very thick] (0,0) -- (0.5,0);})} The difference between the radiated power of a real dipole and the Hertzian dipole follows the relation:
	 \begin{tasks}(4)
		 \task $Hertz = 2\cdot Real$
		 \task $2\cdot Hertz = Real$
		 \task \bm{$Hertz = 4\cdot Real$}
		 \task $4 \cdot Hertz = Real$
		 \end{tasks} 
	 \item \emph{(\protect\tikz[baseline=-0.5ex]{\protect\draw[brown, very thick] (0,0) -- (0.5,0);})} The directivity of a dipole of length $\lambda$/2 is:
	 \begin{tasks}(4)
		 \task 1.64 dBi
		 \task \textbf{2.15 dBi}
		 \task 1.5 dBi
		 \task 1.76 dBi
		 \end{tasks} 
	 \item \emph{(\protect\tikz[baseline=-0.5ex]{\protect\draw[brown, very thick] (0,0) -- (0.5,0);})} For dipoles parallel to a conductive plane, if the distance $h \ll \lambda$, the fields produced by the dipole and its image are:
	 \begin{tasks}(4)
		 \task In phase
		 \task \textbf{In opposite phase}
		 \task Shifted 30 degrees
		 \task Shifted 45 degrees
		 \end{tasks} 
	 \item \emph{(\protect\tikz[baseline=-0.5ex]{\protect\draw[brown, very thick] (0,0) -- (0.5,0);})} The electromotive force is:
	 \begin{tasks}(4)
		 \task It is independent of power
		 \task It depends on power
		 \task \textbf{Independent of distance}
		 \task Dependent on distance
		 \end{tasks} 
	 \item \emph{(\protect\tikz[baseline=-0.5ex]{\protect\draw[brown, very thick] (0,0) -- (0.5,0);})} The directive gain of a $\lambda$/2 dipole is (in linear units):
	 \begin{tasks}(4)
		 \task \textbf{1.64}
		 \task 2.15
		 \task 1.5
		 \task 1.76
		 \end{tasks} 
	 \item \emph{(\protect\tikz[baseline=-0.5ex]{\protect\draw[brown, very thick] (0,0) -- (0.5,0);})} The directivity of a dipole of length $\lambda$ is:
	 \begin{tasks}(4)
		 \task 1.76 dBi
		 \task 2.15 dBi
		 \task \textbf{3.82 dBi}
		 \task 3.36 dBi
		 \end{tasks} 
	 \item \emph{(\protect\tikz[baseline=-0.5ex]{\protect\draw[brown, very thick] (0,0) -- (0.5,0);})} The radiation resistance of a dipole of length 1.5$\lambda$ is:
	 \begin{tasks}(4)
		 \task $36.6 \Omega$
		 \task $73.2 \Omega$
		 \task $\infty$
		 \task \bm{$99.5 \Omega$}
		 \end{tasks} 
	 \item \emph{(\protect\tikz[baseline=-0.5ex]{\protect\draw[brown, very thick] (0,0) -- (0.5,0);})} The directive gain of a short dipole is (in linear units): 
	 \begin{tasks}(4)
		 \task 1.64 
		 \task 2.15
		 \task \textbf{1.5}
		 \task 1.76 
		 \end{tasks} 
	 \item \emph{(\protect\tikz[baseline=-0.5ex]{\protect\draw[brown, very thick] (0,0) -- (0.5,0);})} The difference between the vector potential A of a real dipole and the Hertzian dipole follows the relation:
	 \begin{tasks}(4)
		 \task \bm{$Hertz = 2\cdot Real$}
		 \task $2\cdot Hertz = Real$
		 \task $Hertz = 4\cdot Real$
		 \task $4\cdot Hertz = Real$
		 \end{tasks} 
	 \item \emph{(\protect\tikz[baseline=-0.5ex]{\protect\draw[brown, very thick] (0,0) -- (0.5,0);})} A Short Vertical Antenna (SVA) is a monopole of height $h$ such that:
	 \begin{tasks}(4)
		 \task $h \leq 0.5 \lambda$
		 \task $h \leq \lambda$
		 \task $h \leq 0.3 \lambda$
		 \task \bm{$h \leq 0.1 \lambda$}
		 \end{tasks} 
	 \item \emph{(\protect\tikz[baseline=-0.5ex]{\protect\draw[brown, very thick] (0,0) -- (0.5,0);})} For an elemental dipole over a conductive plane, the total radiated power is \_\_\_\_ compared to free space (for the same current):
	 \begin{tasks}(4)
		 \task Same
		 \task \textbf{Double}
		 \task Triple
		 \task Quadruple
		 \end{tasks} 
	 \item \emph{(\protect\tikz[baseline=-0.5ex]{\protect\draw[brown, very thick] (0,0) -- (0.5,0);})} What is not affected by the presence of the ground in antenna radiation?:
	 \begin{tasks}(4)
		 \task $P_t$
		 \task $\phi(r,\theta,\phi$)
		 \task $R_{rad}$
		 \task \bm{$R_p$}
		 \end{tasks} 
	 \item \emph{(\protect\tikz[baseline=-0.5ex]{\protect\draw[brown, very thick] (0,0) -- (0.5,0);})} The directivity of a dipole of length 1.5$\lambda$ is:
	 \begin{tasks}(4)
		 \task 1.76 dBi
		 \task 2.15 dBi
		 \task 3.82 dBi
		 \task \textbf{3.36 dBi}
		 \end{tasks} 
	 \item \emph{(\protect\tikz[baseline=-0.5ex]{\protect\draw[brown, very thick] (0,0) -- (0.5,0);})} The length of a dipole of length $\lambda$/2 is:
	 \begin{tasks}(4)
		 \task \bm{$L = \lambda/2$} \textbf{(total)}
		 \task $L = \lambda$ (total)
		 \task $L = \lambda/2$ (each arm)
		 \task $L = \lambda/6$ (each arm)
		 \end{tasks} 
	 \item \emph{(\protect\tikz[baseline=-0.5ex]{\protect\draw[brown, very thick] (0,0) -- (0.5,0);})} The difference between the Rrad of a real dipole and the Hertzian dipole follows the relation:
	 \begin{tasks}(4)
		 \task $Hertz = 2\cdot Real$
		 \task $2\cdot Hertz = Real$
		 \task \bm{$Hertz = 4\cdot Real$}
		 \task $4\cdot Hertz = Real$
		 \end{tasks} 
	 \item \emph{(\protect\tikz[baseline=-0.5ex]{\protect\draw[brown, very thick] (0,0) -- (0.5,0);})} The short-circuited transmission line model is used to analyze approximately which antenna?:
	 \begin{tasks}(4)
		 \task Helical antenna
		 \task \textbf{Loop antenna}
		 \task Half-wave dipole
		 \task Yagi–Uda antenna
		 \end{tasks} 
	 \item \emph{(\protect\tikz[baseline=-0.5ex]{\protect\draw[brown, very thick] (0,0) -- (0.5,0);})} The direction of maximum radiation of a $\lambda$/2 dipole is:
	 \begin{tasks}(4)
		 \task \bm{$\theta = \pi/2$}
		 \task $\theta = \pi$
		 \task $\theta = 0$
		 \task $\phi = \pi/2$
		 \end{tasks} 
	 \item \emph{(\protect\tikz[baseline=-0.5ex]{\protect\draw[brown, very thick] (0,0) -- (0.5,0);})} The radiation pattern in power of a short dipole is:
	 \begin{tasks}(4)
		 \task \bm{$f(\theta,\phi) = \sin^2(\theta)$}
		 \task $f(\theta,\phi) = \cos^2(\theta)$
		 \task $f(\theta,\phi) = \sin^2(\phi)$
		 \task $f(\theta,\phi) = \cos^2(\phi)$
		 \end{tasks} 
	 \item \emph{(\protect\tikz[baseline=-0.5ex]{\protect\draw[brown, very thick] (0,0) -- (0.5,0);})} For a dipole parallel to a conductive plane and separated by $\lambda$/4, the total field is \_\_\_\_ compared to free space (for the same current):
	 \begin{tasks}(4)
		 \task Same
		 \task \textbf{Double}
		 \task Triple
		 \task Quadruple
		 \end{tasks} 
	 \item \emph{(\protect\tikz[baseline=-0.5ex]{\protect\draw[brown, very thick] (0,0) -- (0.5,0);})} The radiation resistance of a dipole at $\lambda$/2 is:
	 \begin{tasks}(4)
		 \task $1.64 \Omega$
		 \task $36.2 \Omega$
		 \task \bm{$73.2 \Omega$}
		 \task $50 \Omega$
		 \end{tasks} 
	 \item \emph{(\protect\tikz[baseline=-0.5ex]{\protect\draw[brown, very thick] (0,0) -- (0.5,0);})} By the image theorem, the elemental monopole over a conductive plane is equivalent to a \_\_\_\_ fed with the same current in free space:
	 \begin{tasks}(4)
		 \task \textbf{Double-length dipole}
		 \task Same-length dipole
		 \task Double-length monopole
		 \task Same-length monopole
		 \end{tasks} 
	 \item \emph{(\protect\tikz[baseline=-0.5ex]{\protect\draw[brown, very thick] (0,0) -- (0.5,0);})} The antennas placed over a conductor plane are analyzed through the:.
	 \begin{tasks}(4)
		 \task Method of reflections
		 \task Method of reflectivity
		 \task Maxwell's method
		 \task \textbf{Method of images}
		 \end{tasks} 
	 \item \emph{(\protect\tikz[baseline=-0.5ex]{\protect\draw[brown, very thick] (0,0) -- (0.5,0);})} For an isotropic antenna over a conductive plane (c.p.) radiating Pt, what is the field relation compared to its free-space (f.s.) equivalent?: (c.p. = conductive plane; f.s. = free space; Pt = transmitted (radiated) power; $\phi$ = Power flux density) 
	 \begin{tasks}(4)
		 \task $\phi_{c.p.} = 2\cdot \phi_{f.s.}$
		 \task $2\cdot \phi_{c.p.} = \phi_{f.s.}$
		 \task \bm{$\phi_{c.p.} = \sqrt{2} \cdot \phi_{f.s.}$}
		 \task $\sqrt{2} \cdot \phi_{c.p..} = \phi_{f.s.}$
		 \end{tasks} 
	 \item \emph{(\protect\tikz[baseline=-0.5ex]{\protect\draw[brown, very thick] (0,0) -- (0.5,0);})} The apparent radiated power of a monopole with respect to the SVA is known by the acronym:
	 \begin{tasks}(4)
		 \task EIRP
		 \task EMEP
		 \task EMRA
		 \task \textbf{EMRP}
		 \end{tasks} 
	 \item \emph{(\protect\tikz[baseline=-0.5ex]{\protect\draw[brown, very thick] (0,0) -- (0.5,0);})} For dipoles parallel to a conductive plane, if the distance $h = \lambda/4$, the fields produced by the dipole and its image are:
	 \begin{tasks}(4)
		 \task \textbf{In phase}
		 \task In opposite phase
		 \task Shifted 30 degrees
		 \task Shifted 45 degrees
		 \end{tasks} 
	 \item \emph{(\protect\tikz[baseline=-0.5ex]{\protect\draw[orange, very thick] (0,0) -- (0.5,0);})} The differential solid angle can be expressed in spherical coordinates as:
	 \begin{tasks}(4)
		 \task $\cos\theta d\theta d\phi$
		 \task \bm{$\sin\theta d\theta d\phi$}
		 \task $\cos\phi d\theta d\phi$
		 \task $\sin\phi d\theta d\phi$
		 \end{tasks} 
	 \item \emph{(\protect\tikz[baseline=-0.5ex]{\protect\draw[orange, very thick] (0,0) -- (0.5,0);})} The magnitude of $\vec{E}$ and $\vec{H}$ radiated by a current element at (0,0) aligned with $z$ varies with the angular coordinates as:
	 \begin{tasks}(4)
		 \task $\cos\theta$
		 \task \bm{$\sin\theta$}
		 \task $\sin\phi$
		 \task $\sin\theta \cos\phi$
		 \end{tasks} 
	 \item \emph{(\protect\tikz[baseline=-0.5ex]{\protect\draw[orange, very thick] (0,0) -- (0.5,0);})} One of the expressions of $EIRP(\theta,\phi)$ is: 
	 \begin{tasks}(4)
		 \task $P_t \cdot g(\theta,\phi)$
		 \task \bm{$P_t \cdot d(\theta,\phi)$}
		 \task $P_t \cdot EIRP$
		 \task $L_t \cdot g(\theta,\phi)$
		 \end{tasks} 
	 \item \emph{(\protect\tikz[baseline=-0.5ex]{\protect\draw[orange, very thick] (0,0) -- (0.5,0);})} Directivity in logarithmic units is measured in dBi when:
	 \begin{tasks}(4)
		 \task Always
		 \task When the value is $>$ 1
		 \task It has no value, it is a figure of merit
		 \task \textbf{When the reference antenna is isotropic}
		 \end{tasks} 
	 \item \emph{(\protect\tikz[baseline=-0.5ex]{\protect\draw[orange, very thick] (0,0) -- (0.5,0);})} Sector antennas are:
	 \begin{tasks}(4)
		 \task Pencil-beam ones
		 \task Omnidirectional ones
		 \task \textbf{Fan-beam ones}
		 \task Contoured-beam ones
		 \end{tasks} 
	 \item \emph{(\protect\tikz[baseline=-0.5ex]{\protect\draw[orange, very thick] (0,0) -- (0.5,0);})} To facilitate the calculation of the field radiated by an arbitrary source, the distance $|\vec{r}-\vec{r'}|$ is approximated in the far-field region by: 
\begin{tasks}(4)
    \task $\vec{r} - \hat{r} v'$
    \task $\vec{r} - r v'$
    \task $\vec{r} - r v'$
    \task \bm{$r - \hat{r}\cdot \vec{r'}$}
\end{tasks}
	 \item \emph{(\protect\tikz[baseline=-0.5ex]{\protect\draw[orange, very thick] (0,0) -- (0.5,0);})} The effective aperture of a receiving antenna is measured in:
	 \begin{tasks}(4)
		 \task It has no units
		 \task \bm{$\mathrm{m}^2$}
		 \task W
		 \task $\mathrm{m}^{-2}$
		 \end{tasks} 
	 \item \emph{(\protect\tikz[baseline=-0.5ex]{\protect\draw[orange, very thick] (0,0) -- (0.5,0);})} The power flux density radiated per unit surface area is: 
	 \begin{tasks}(4)
		 \task It does not depend on angular units
		 \task \textbf{Measured in  W/m$^2$}
		 \task Measured in W/Solid angle
		 \task Independent of distance
		 \end{tasks} 
	 \item \emph{(\protect\tikz[baseline=-0.5ex]{\protect\draw[orange, very thick] (0,0) -- (0.5,0);})} To model an antenna in the noise system, we use a:
	 \begin{tasks}(4)
		 \task \textbf{Dipole}
		 \task Monopole
		 \task Quadrupole
		 \task Pentapole
		 \end{tasks} 
	 \item \emph{(\protect\tikz[baseline=-0.5ex]{\protect\draw[orange, very thick] (0,0) -- (0.5,0);})} The directivity approximation that follows the formula $4\pi/(BW_{-3dB}\cdot BW_{-3dB})$ is called:
	 \begin{tasks}(4)
		 \task Sanmartín approximation
		 \task Gauss approximation
		 \task Tai approximation
		 \task \textbf{Krauss approximation}
		 \end{tasks} 
	 \item \emph{(\protect\tikz[baseline=-0.5ex]{\protect\draw[orange, very thick] (0,0) -- (0.5,0);})} Noise limits:
	 \begin{tasks}(4)
		 \task Antenna temperature
		 \task Received signal
		 \task \textbf{Quality of a radio system}
		 \task Bandwidth
		 \end{tasks} 
	 \item \emph{(\protect\tikz[baseline=-0.5ex]{\protect\draw[orange, very thick] (0,0) -- (0.5,0);})} The antenna that extracts all the power from the medium and delivers it entirely to a matched receiver is called:
	 \begin{tasks}(4)
		 \task Special
		 \task Directive
		 \task Franklin
		 \task \textbf{Ideal}
		 \end{tasks} 
	 \item \emph{(\protect\tikz[baseline=-0.5ex]{\protect\draw[orange, very thick] (0,0) -- (0.5,0);})} The carrier-to-interference ratio and protection ratio are equivalent except for the term:
	 \begin{tasks}(4)
		 \task Antenna powers
		 \task Antenna distances
		 \task Antenna size
		 \task \textbf{Antenna directivities}
		 \end{tasks} 
	 \item \emph{(\protect\tikz[baseline=-0.5ex]{\protect\draw[orange, very thick] (0,0) -- (0.5,0);})} The retarded potentials $\vec{A}$ and $V$ are retarded because:
	 \begin{tasks}(4)
		 \task They are dumb
		 \task They have no physical meaning
		 \task \textbf{In Helmholtz equations, there is a time shift}
		 \task They came after Maxwell’s equations
		 \end{tasks} 
	 \item \emph{(\protect\tikz[baseline=-0.5ex]{\protect\draw[orange, very thick] (0,0) -- (0.5,0);})} The retarded potentials $\vec{A}$ and $V$:
	 \begin{tasks}(4)
		 \task They propagate as inhomogeneous waves
		 \task Related by the divergence condition
		 \task They do not have the same value at all points of the wavefront
		 \task \bm{$\vec{A}$} \textbf{is a vector and} \bm{$V$} \textbf{is a scalar}
		 \end{tasks} 
	 \item \emph{(\protect\tikz[baseline=-0.5ex]{\protect\draw[orange, very thick] (0,0) -- (0.5,0);})} According to the distance from the antenna, which zone does not exist when classifying space:
	 \begin{tasks}(4)
		 \task Radiating near field
		 \task \textbf{Super field}
		 \task Reactive near field
		 \task Fraunhofer zone
		 \end{tasks} 
	 \item \emph{(\protect\tikz[baseline=-0.5ex]{\protect\draw[orange, very thick] (0,0) -- (0.5,0);})} Power gain follows the same principle as directivity but takes into account:
	 \begin{tasks}(4)
		 \task \textbf{Delivered power}
		 \task Average power
		 \task Transmitted power
		 \task Radiated power
		 \end{tasks} 
	 \item \emph{(\protect\tikz[baseline=-0.5ex]{\protect\draw[orange, very thick] (0,0) -- (0.5,0);})} The direction cosines ($u,v$) determine:
	 \begin{tasks}(4)
		 \task Position in the imaginary plane
		 \task Position on a unit sphere
		 \task \textbf{A direction in space}
		 \task A direction in the vertical plane
		 \end{tasks} 
	 \item \emph{(\protect\tikz[baseline=-0.5ex]{\protect\draw[orange, very thick] (0,0) -- (0.5,0);})} The concept of interference gives rise to two different calculations:
	 \begin{tasks}(4)
		 \task \textbf{Carrier-to-Interference ratio and protection ratio}
		 \task Field relations and protection ratio
		 \task Protection ratio and interfering ratio
		 \task Co-channel and adjacent ratio
		 \end{tasks} 
	 \item \emph{(\protect\tikz[baseline=-0.5ex]{\protect\draw[orange, very thick] (0,0) -- (0.5,0);})} The power dissipated by Joule effect in an antenna is characterized by:
	 \begin{tasks}(4)
		 \task $R_{rad}$
		 \task Loss tangent
		 \task \bm{$R_{losses}$}
		 \task $R_{reactance}$
		 \end{tasks} 
	 \item \emph{(\protect\tikz[baseline=-0.5ex]{\protect\draw[orange, very thick] (0,0) -- (0.5,0);})} The Helmholtz equation is:
	 \begin{tasks}(4)
		 \task \textbf{A wave equation}
		 \task One of Maxwell’s equations
		 \task An equation linking retarded potentials
		 \task The solution of a Bessel equation
		 \end{tasks} 
	 \item \emph{(\protect\tikz[baseline=-0.5ex]{\protect\draw[orange, very thick] (0,0) -- (0.5,0);})} The basic free-space propagation loss depends on:
	 \begin{tasks}(4)
		 \task The antennas used
		 \task Obstacles in the path
		 \task Atmospheric phenomena
		 \task \textbf{Frequency and distance}
		 \end{tasks} 
	 \item \emph{(\protect\tikz[baseline=-0.5ex]{\protect\draw[orange, very thick] (0,0) -- (0.5,0);})} The ratio of maximum to minimum of the operating band of an antenna with a bandwidth of one octave is:
	 \begin{tasks}(4)
		 \task 8 : 1
		 \task \textbf{2 : 1}
		 \task 8 : 2
		 \task 2 : 2
		 \end{tasks} 
	 \item \emph{(\protect\tikz[baseline=-0.5ex]{\protect\draw[orange, very thick] (0,0) -- (0.5,0);})} In the far field: the electric and magnetic fields produced by a current element at (0,0) aligned with $z$:
	 \begin{tasks}(4)
		 \task \textbf{Locally behaves like a plane wave}
		 \task Constitutes a homogeneous spherical wave
		 \task Both fields are parallel to each other
		 \task The ratio H/E is called intrinsic impedance of free space
		 \end{tasks} 
	 \item \emph{(\protect\tikz[baseline=-0.5ex]{\protect\draw[orange, very thick] (0,0) -- (0.5,0);})} The Helmholtz equations allow deducing that:
	 \begin{tasks}(4)
		 \task The vector potential $\vec{A}$ is related to charge density
		 \task Scalar potential $V$ is related to current density
		 \task \textbf{Potentials are related to each other}
		 \task Both $\vec{A}$ and $V$ are related to charge densities
		 \end{tasks} 
	 \item \emph{(\protect\tikz[baseline=-0.5ex]{\protect\draw[orange, very thick] (0,0) -- (0.5,0);})} The loss tangent has a greater value when: 
	 \begin{tasks}(4)
		 \task The losses are smaller
		 \task Higher frequency
		 \task \textbf{Higher material conductivity}
		 \task Greater angle of wave incidence
		 \end{tasks} 
	 \item \emph{(\protect\tikz[baseline=-0.5ex]{\protect\draw[orange, very thick] (0,0) -- (0.5,0);})} The solid angle is denoted by the Greek letter: 
	 \begin{tasks}(4)
		 \task \bm{$\Omega$}
		 \task $\Gamma$
		 \task $\gamma$
		 \task $\theta$
		 \end{tasks} 
	 \item \emph{(\protect\tikz[baseline=-0.5ex]{\protect\draw[orange, very thick] (0,0) -- (0.5,0);})} The noise figure of an amplifier is:
	 \begin{tasks}(4)
		 \task The same as noise temperature
		 \task The physical noise of the antenna
		 \task \textbf{A figure of merit}
		 \task Never at $T_0$
		 \end{tasks} 
	 \item \emph{(\protect\tikz[baseline=-0.5ex]{\protect\draw[orange, very thick] (0,0) -- (0.5,0);})} The shape of the radiation pattern of omnidirectional antennas usually looks like:
	 \begin{tasks}(4)
		 \task A cylinder
		 \task A cone
		 \task A sphere
		 \task \textbf{A “toroid”}
		 \end{tasks} 
	 \item \emph{(\protect\tikz[baseline=-0.5ex]{\protect\draw[orange, very thick] (0,0) -- (0.5,0);})} The co-channel electrical field protection ratio is a measure of the intensity difference between: 
	 \begin{tasks}(4)
		 \task Desired and interfering power
		 \task \textbf{Desired and interfering E-field strength}
		 \task Received and transmitted power
		 \task E-field of 2 interfering fields
		 \end{tasks} 
	 \item \emph{(\protect\tikz[baseline=-0.5ex]{\protect\draw[orange, very thick] (0,0) -- (0.5,0);})} The radiation resistance of an antenna is:
	 \begin{tasks}(4)
		 \task It should preferably be low
		 \task The input resistance the antenna offers when transmitting
		 \task Independent of frequency
		 \task \textbf{Represents the antenna’s ability to convert delivered power into $P_{rad}$}
		 \end{tasks} 
	 \item \emph{(\protect\tikz[baseline=-0.5ex]{\protect\draw[orange, very thick] (0,0) -- (0.5,0);})} The power gain:
	 \begin{tasks}(4)
		 \task Does not take antenna losses into account
		 \task \textbf{Depends on antenna efficiency}
		 \task Varies quite a lot from directive gain
		 \task Takes into account impedance mismatch
		 \end{tasks} 
	 \item \emph{(\protect\tikz[baseline=-0.5ex]{\protect\draw[orange, very thick] (0,0) -- (0.5,0);})} The antenna efficiency can be calculated with which relation?:
	 \begin{tasks}(4)
		 \task $2\cdot (P_e / P_t)$
		 \task $P_e / P_t$
		 \task \bm{$P_t/ P_e$}
		 \task $2\cdot (P_t / P_e)$
		 \end{tasks} 
	 \item \emph{(\protect\tikz[baseline=-0.5ex]{\protect\draw[orange, very thick] (0,0) -- (0.5,0);})} What does $proj_u \vec{r}$ mean mathematically?: 
	 \begin{tasks}(4)
		 \task An “$\vec{r}$” that is cold
		 \task \textbf{The projection of vector $\vec{r}$}
		 \task The vector perpendicular to $\vec{r}$
		 \task Scalar value of vector parallel to $\vec{r}$
		 \end{tasks} 
	 \item \emph{(\protect\tikz[baseline=-0.5ex]{\protect\draw[orange, very thick] (0,0) -- (0.5,0);})} Field attenuation is introduced to take into account the influence on the basic propagation loss of:
	 \begin{tasks}(4)
		 \task Distance
		 \task \textbf{The propagation medium}
		 \task Interference from other sources
		 \task The operating frequency
		 \end{tasks} 
	 \item \emph{(\protect\tikz[baseline=-0.5ex]{\protect\draw[orange, very thick] (0,0) -- (0.5,0);})} The side lobes of a radiation pattern are:
	 \begin{tasks}(4)
		 \task \textbf{Those that are not the main one(s)}
		 \task The main one(s)
		 \task Those opposite to the main direction
		 \task Those perpendicular to the main direction
		 \end{tasks} 
	 \item \emph{(\protect\tikz[baseline=-0.5ex]{\protect\draw[orange, very thick] (0,0) -- (0.5,0);})} The power flux density radiated by an isotropic antenna is:
	 \begin{tasks}(4)
		 \task \bm{$P_t/(4\pi r^2)$}
		 \task $P_t/ (4\pi)$
		 \task $P_t/ (2\pi r)$
		 \task $P_t/ (\pi r^2)$
		 \end{tasks} 
	 \item \emph{(\protect\tikz[baseline=-0.5ex]{\protect\draw[orange, very thick] (0,0) -- (0.5,0);})} The radiation intensity of the isotropic antenna is:
	 \begin{tasks}(4)
		 \task $P_t/ 4\pi r^2$
		 \task \bm{$P_t/ 4\pi$}
		 \task $P_t/ 2\pi r$
		 \task $P_t/ \pi r^2$
		 \end{tasks} 
	 \item \emph{(\protect\tikz[baseline=-0.5ex]{\protect\draw[orange, very thick] (0,0) -- (0.5,0);})} The number of steradians in a sphere is:
	 \begin{tasks}(4)
		 \task $2\pi$
		 \task $(4/3)\pi r^3$
		 \task $4\pi r$
		 \task \textbf{$4\pi$}
		 \end{tasks} 
	 \item \emph{(\protect\tikz[baseline=-0.5ex]{\protect\draw[orange, very thick] (0,0) -- (0.5,0);})} Directive gain is defined as:
	 \begin{tasks}(4)
		 \task $g(\theta,\phi) = e(\theta,\phi) / e_{iso}$
		 \task $g(\theta,\phi) = \phi(\theta,\phi) / \phi_{iso}$
		 \task $g(\theta,\phi) = i_{max} / i_{iso}$
		 \task \bm{$g(\theta,\phi) = i(\theta,\phi)/i_{iso}$}
		 \end{tasks} 
	 \item \emph{(\protect\tikz[baseline=-0.5ex]{\protect\draw[orange, very thick] (0,0) -- (0.5,0);})} Radio wave propagation in real life behaves as:
	 \begin{tasks}(4)
		 \task Jacobian functions
		 \task Exact nominal values
		 \task Bessel functions
		 \task \textbf{Statistical distribution}
		 \end{tasks} 
	 \item \emph{(\protect\tikz[baseline=-0.5ex]{\protect\draw[orange, very thick] (0,0) -- (0.5,0);})} The reflection coefficient, VSWR, and return loss measure:
	 \begin{tasks}(4)
		 \task The transmitter–receiver relationship
		 \task \textbf{Mismatch}
		 \task Impedance
		 \task Power
		 \end{tasks} 
	 \item \emph{(\protect\tikz[baseline=-0.5ex]{\protect\draw[orange, very thick] (0,0) -- (0.5,0);})} For right-hand circular polarization, it is called: 
	 \begin{tasks}(4)
		 \task \textbf{Dextrorotatory}
		 \task Dextrory
		 \task Levory
		 \task Levorotatory
		 \end{tasks} 
	 \item \emph{(\protect\tikz[baseline=-0.5ex]{\protect\draw[orange, very thick] (0,0) -- (0.5,0);})} For left-hand circular polarization, it is called:
	 \begin{tasks}(4)
		 \task Dextrorotatory
		 \task Dextrory
		 \task Levory
		 \task \textbf{Levorotatory}
		 \end{tasks} 
	 \item \emph{(\protect\tikz[baseline=-0.5ex]{\protect\draw[orange, very thick] (0,0) -- (0.5,0);})} The power radiated per unit solid angle is called:
	 \begin{tasks}(4)
		 \task Directivity
		 \task Power flux density
		 \task \textbf{Radiation intensity}
		 \task Radiation resistance
		 \end{tasks} 
	 \item \emph{(\protect\tikz[baseline=-0.5ex]{\protect\draw[orange, very thick] (0,0) -- (0.5,0);})} The integrals used to solve directivity problems are called Euler integrals, and their notation was formalized by:
	 \begin{tasks}(4)
		 \task Fourier
		 \task \textbf{Legendre}
		 \task Euler
		 \task Mika
		 \end{tasks} 
	 \item \emph{(\protect\tikz[baseline=-0.5ex]{\protect\draw[orange, very thick] (0,0) -- (0.5,0);})} The (maximum) directivity of an antenna: 
	 \begin{tasks}(4)
		 \task Depends on the angular variables
		 \task Cannot be $>$ 1 dB in real antennas
		 \task Is the power radiated
		 \task \textbf{The value 1 (or 0 dBi) is only valid for isotropic antennas}
		 \end{tasks} 
	 \item \emph{(\protect\tikz[baseline=-0.5ex]{\protect\draw[orange, very thick] (0,0) -- (0.5,0);})} An isotropic antenna is:
	 \begin{tasks}(4)
		 \task It is a tropical antenna
		 \task It is a satellite antenna
		 \task Equivalent to a current element
		 \task \textbf{It is an abstraction. Not physically realizable}
		 \end{tasks} 
	 \item \emph{(\protect\tikz[baseline=-0.5ex]{\protect\draw[green, very thick] (0,0) -- (0.5,0);})} Given “a” as the width and “b” as the height of a rectangular waveguide, a sectoral H-plane horn would have which dimensions?
	 \begin{tasks}(4)
		 \task A $\times$ B
		 \task \textbf{A $\times$ b}
		 \task B $\times$ a
		 \task a $\times$ b
		 \end{tasks} 
	 \item \emph{(\protect\tikz[baseline=-0.5ex]{\protect\draw[green, very thick] (0,0) -- (0.5,0);})} The H-plane contains the direction of energy radiation and:
	 \begin{tasks}(4)
		 \task Electric field
		 \task \textbf{Magnetic field}
		 \task Electromagnetic field
		 \task Quantum field
		 \end{tasks} 
	 \item \emph{(\protect\tikz[baseline=-0.5ex]{\protect\draw[green, very thick] (0,0) -- (0.5,0);})} A broadside array has its maximum radiation direction at:
	 \begin{tasks}(4)
		 \task \bm{$\alpha = 0$}
		 \task $\alpha = -k_0 d$
		 \task $\alpha = +k_0 d$
		 \task Variable $\alpha$
		 \end{tasks} 
	 \item \emph{(\protect\tikz[baseline=-0.5ex]{\protect\draw[green, very thick] (0,0) -- (0.5,0);})} Given “a” as the width and “b” as the height of a rectangular waveguide, a sectoral E-plane horn would have which dimensions?
	 \begin{tasks}(4)
		 \task A $\times$ B
		 \task A $\times$ b
		 \task \textbf{B $\times$ a}
		 \task b $\times$ a
		 \end{tasks} 
	 \item \emph{(\protect\tikz[baseline=-0.5ex]{\protect\draw[green, very thick] (0,0) -- (0.5,0);})} The aperture illumination efficiency gives an idea of:
	 \begin{tasks}(4)
		 \task \textbf{How uniform its illumination field is}
		 \task What spills at the dish edges
		 \task Diffraction
		 \task All of the above
		 \end{tasks} 
	 \item \emph{(\protect\tikz[baseline=-0.5ex]{\protect\draw[green, very thick] (0,0) -- (0.5,0);})} Which of the following antennas IS an aperture antenna?
	 \begin{tasks}(4)
		 \task Dipole
		 \task Franklin
		 \task \textbf{Gregorian}
		 \task Microstrip patch
		 \end{tasks} 
	 \item \emph{(\protect\tikz[baseline=-0.5ex]{\protect\draw[green, very thick] (0,0) -- (0.5,0);})} When the diameter or size of an aperture increases, its main beamwidth:
	 \begin{tasks}(4)
		 \task Increases
		 \task \textbf{Decreases}
		 \task Depends
		 \task None of the above
		 \end{tasks} 
	 \item \emph{(\protect\tikz[baseline=-0.5ex]{\protect\draw[green, very thick] (0,0) -- (0.5,0);})} Increasing the aperture size increases:
	 \begin{tasks}(4)
		 \task \textbf{Directivity}
		 \task Efficiency
		 \task Cutoff frequency in a waveguide
		 \task Our patience
		 \end{tasks} 
	 \item \emph{(\protect\tikz[baseline=-0.5ex]{\protect\draw[green, very thick] (0,0) -- (0.5,0);})} A TE mode in a waveguide that extends along the $z$ direction satisfies:
	 \begin{tasks}(4)
		 \task \bm{$E_z = 0$}
		 \task $H_z = 0$
		 \task $E_z = H_z = 0$
		 \task None of the above
		 \end{tasks} 
	 \item \emph{(\protect\tikz[baseline=-0.5ex]{\protect\draw[green, very thick] (0,0) -- (0.5,0);})} The fundamental mode of a rectangular waveguide is:
	 \begin{tasks}(4)
		 \task TM$_{10}$
		 \task \bm{$\mathrm{TE}_{10}$}
		 \task TM$_{00}$
		 \task TE$_{00}$
		 \end{tasks} 
	 \item \emph{(\protect\tikz[baseline=-0.5ex]{\protect\draw[green, very thick] (0,0) -- (0.5,0);})} A reflector fed by a horn with linear polarization has (in the ideal case):
	 \begin{tasks}(4)
		 \task \textbf{Linear polarization}
		 \task Circular polarization
		 \task Elliptical polarization
		 \task Solar polarization
		 \end{tasks} 
	 \item \emph{(\protect\tikz[baseline=-0.5ex]{\protect\draw[green, very thick] (0,0) -- (0.5,0);})} The parameter “$f$” in a reflector indicates:
	 \begin{tasks}(4)
		 \task The reflector diameter
		 \task Maximum distance
		 \task Aperture angle
		 \task \textbf{Focal distance}
		 \end{tasks} 
	 \item \emph{(\protect\tikz[baseline=-0.5ex]{\protect\draw[green, very thick] (0,0) -- (0.5,0);})} The tapering losses in a parabolic reflector are associated with:
	 \begin{tasks}(4)
		 \task Radiation that does not hit the dish
		 \task Ohmic losses
		 \task \textbf{Non-uniform feed}
		 \task Metal roughness
		 \end{tasks} 
	 \item \emph{(\protect\tikz[baseline=-0.5ex]{\protect\draw[green, very thick] (0,0) -- (0.5,0);})} In practice, the “$f/D$” ratio usually is in the range:
	 \begin{tasks}(4)
		 \task (0.1, 0.9)
		 \task \textbf{(0.3, 0.4)}
		 \task (0.8, 0.9)
		 \task 2
		 \end{tasks} 
	 \item \emph{(\protect\tikz[baseline=-0.5ex]{\protect\draw[green, very thick] (0,0) -- (0.5,0);})} The spillover losses are associated with:
	 \begin{tasks}(4)
		 \task Ohmic losses
		 \task Feed losses
		 \task \textbf{Radiation that does not hit the dish}
		 \task Metal roughness
		 \end{tasks} 
	 \item \emph{(\protect\tikz[baseline=-0.5ex]{\protect\draw[green, very thick] (0,0) -- (0.5,0);})} A conical horn is the natural extension of a waveguide:
	 \begin{tasks}(4)
		 \task Pyramidal
		 \task \textbf{Circular}
		 \task H-plane sectorial
		 \task Optical fiber
		 \end{tasks} 
	 \item \emph{(\protect\tikz[baseline=-0.5ex]{\protect\draw[green, very thick] (0,0) -- (0.5,0);})} The Array Factor for simple, uniformly spaced linear arrays with uniform excitation and no element tapering has the form of: 
	 \begin{tasks}(4)
		 \task Sine
		 \task Cosine
		 \task Tangent
		 \task \textbf{sinc}
		 \end{tasks} 
	 \item \emph{(\protect\tikz[baseline=-0.5ex]{\protect\draw[green, very thick] (0,0) -- (0.5,0);})} A reflector antenna is composed of:
	 \begin{tasks}(4)
		 \task Receiving antenna $+$ metal reflector
		 \task Receiving antenna $+$ non-metallic reflector
		 \task \textbf{Feed antenna $+$ metallic reflector}
		 \task Feed antenna $+$ non-metallic reflector
		 \end{tasks} 
	 \item \emph{(\protect\tikz[baseline=-0.5ex]{\protect\draw[green, very thick] (0,0) -- (0.5,0);})} An Endfire array has its maximum radiation direction at:
	 \begin{tasks}(4)
		 \task $\alpha = 0$
		 \task $\bm{\alpha = -k_0 d}$
		 \task $\alpha = +k_0 d$
		 \task Variable $ \alpha$
		 \end{tasks} 
	 \item \emph{(\protect\tikz[baseline=-0.5ex]{\protect\draw[green, very thick] (0,0) -- (0.5,0);})} A hybrid mode in a waveguide or horn that extends along the $z$ direction fulfills:
	 \begin{tasks}(4)
		 \task $E_z = 0$
		 \task $H_z = 0$
		 \task TEM
		 \task \bm{$E_z \neq 0; H_z \neq 0$}
		 \end{tasks} 
	 \item \emph{(\protect\tikz[baseline=-0.5ex]{\protect\draw[green, very thick] (0,0) -- (0.5,0);})} The element that concentrates the rays at a point called the focus is:
	 \begin{tasks}(4)
		 \task Line
		 \task \textbf{Parabola}
		 \task Circle
		 \task Both of the above
		 \end{tasks} 
	 \item \emph{(\protect\tikz[baseline=-0.5ex]{\protect\draw[green, very thick] (0,0) -- (0.5,0);})} In a progressive phase array with alpha, it is used for:
	 \begin{tasks}(4)
		 \task Keep the direction of maximum propagation fixed
		 \task Establish the array current
		 \task \textbf{Change the direction of maximum propagation electronically}
		 \task Change the direction of maximum propagation mechanically
		 \end{tasks} 
	 \item \emph{(\protect\tikz[baseline=-0.5ex]{\protect\draw[green, very thick] (0,0) -- (0.5,0);})} In aperture antenna analysis, we use:
	 \begin{tasks}(4)
		 \task Principle of Action
		 \task The Little Prince principle
		 \task \textbf{Huygens’ principle}
		 \task Principle of the end
		 \end{tasks} 
	 \item \emph{(\protect\tikz[baseline=-0.5ex]{\protect\draw[green, very thick] (0,0) -- (0.5,0);})} In array factor calculations for a linear array, $\Psi$ represents the phase difference between adjacent elements. In the following expression, $\Psi = kd \cos\theta + \alpha$, the angle of observation corresponds to: 
	 \begin{tasks}(4)
		 \task $\phi$
		 \task \bm{$\theta$}
		 \task $\alpha$
		 \task $\beta$
		 \end{tasks} 
	 \item \emph{(\protect\tikz[baseline=-0.5ex]{\protect\draw[green, very thick] (0,0) -- (0.5,0);})} In an array, each antenna is called:
	 \begin{tasks}(4)
		 \task Elemental
		 \task Tiny antenna
		 \task \textbf{Element}
		 \task Antenna
		 \end{tasks} 
	 \item \emph{(\protect\tikz[baseline=-0.5ex]{\protect\draw[green, very thick] (0,0) -- (0.5,0);})} To increase the directivity of an array, we can:
	 \begin{tasks}(4)
		 \task Decrease the number of antennas ($N$)
		 \task \textbf{Increase the number of antennas ($N$)}
		 \task Transmit more power
		 \task Put a mirror
		 \end{tasks} 
	 \item \emph{(\protect\tikz[baseline=-0.5ex]{\protect\draw[green, very thick] (0,0) -- (0.5,0);})} A scanning array has its maximum radiation direction at:
	 \begin{tasks}(4)
		 \task $\alpha = 0$
		 \task $\alpha =-k_0 d$
		 \task $\alpha = +k_0 d$
		 \task \bm{\mathrm{Variable } $\alpha$}
		 \end{tasks} 
	 \item \emph{(\protect\tikz[baseline=-0.5ex]{\protect\draw[green, very thick] (0,0) -- (0.5,0);})} The following structure is NOT a waveguide:
	 \begin{tasks}(4)
		 \task Optical fiber
		 \task \textbf{Coaxial}
		 \task Circular waveguide
		 \task Rectangular waveguide
		 \end{tasks} 
	 \item \emph{(\protect\tikz[baseline=-0.5ex]{\protect\draw[pink, very thick] (0,0) -- (0.5,0);})} Scintillation is produced by:
	 \begin{tasks}(4)
		 \task \textbf{Irregularities in the troposphere}
		 \task Irregularities on the Earth’s surface
		 \task Ionospheric irregularities
		 \task Radio link quality
		 \end{tasks} 
	 \item \emph{(\protect\tikz[baseline=-0.5ex]{\protect\draw[pink, very thick] (0,0) -- (0.5,0);})} 
In the fundamental propagation equation (neglecting the Surface Wave), the normalized field $\frac{|e|}{e_0}$ behaves as a function that is: 
\begin{tasks}(4)
    \task \textbf{Periodic in the inverse of $d$}
    \task Non-periodic in the inverse of $d$
    \task Periodic in the inverse of $f$ 
    \task Non-periodic in the inverse of $f$
\end{tasks} 
	 
	 \item \emph{(\protect\tikz[baseline=-0.5ex]{\protect\draw[pink, very thick] (0,0) -- (0.5,0);})} If the effective Earth radius factor ($k$) is greater than 1:
	 \begin{tasks}(4)
		 \task \bm{$R_{(eff.)} > R_{(real)}$}
		 \task $R_{(eff.)} < R_{(real)}$
		 \task $R_{(eff.)} = R_{(real)}$
		 \task $R_{(eff.)} = \sqrt{R_{(real)}}$
		 \end{tasks} 
	 \item \emph{(\protect\tikz[baseline=-0.5ex]{\protect\draw[pink, very thick] (0,0) -- (0.5,0);})} The three types of “specific attenuation” to be considered are:
	 \begin{tasks}(4)
		 \task Vegetation, atmospheric gases, and fog
		 \task Reflection, atmospheric gases, and rain
		 \task \textbf{Vegetation, atmospheric gases, and rain}
		 \task Vegetation, Earth shine, and rain
		 \end{tasks} 
	 \item \emph{(\protect\tikz[baseline=-0.5ex]{\protect\draw[pink, very thick] (0,0) -- (0.5,0);})} The most dangerous type of fading is:
	 \begin{tasks}(4)
		 \task Scintillation
		 \task Of power
		 \task Pearson
		 \task \textbf{Multipath}
		 \end{tasks} 
	 \item \emph{(\protect\tikz[baseline=-0.5ex]{\protect\draw[pink, very thick] (0,0) -- (0.5,0);})} The methods for calculating fading probability such as Mojoli or P.530 are applied to:
	 \begin{tasks}(4)
		 \task The most unfavorable day
		 \task \textbf{The most unfavorable month}
		 \task The most favorable month
		 \task The most unfavorable year
		 \end{tasks} 
	 \item \emph{(\protect\tikz[baseline=-0.5ex]{\protect\draw[pink, very thick] (0,0) -- (0.5,0);})} The antenna used in Surface Wave propagation is:
	 \begin{tasks}(4)
		 \task Yagi–Uda
		 \task Dipole facing a plane
		 \task Elric antenna
		 \task \textbf{Vertical monopole}
		 \end{tasks} 
	 \item \emph{(\protect\tikz[baseline=-0.5ex]{\protect\draw[pink, very thick] (0,0) -- (0.5,0);})} Curvature is:
	 \begin{tasks}(4)
		 \task The radius of a planet
		 \task The Earth’s radius
		 \task Directly proportional to the radius
		 \task \textbf{Inversely proportional to radius}
		 \end{tasks} 
	 \item \emph{(\protect\tikz[baseline=-0.5ex]{\protect\draw[pink, very thick] (0,0) -- (0.5,0);})} We differentiate different modes of electromagnetic wave propagation depending on:
	 \begin{tasks}(4)
		 \task The link distance
		 \task The type of conductor that guides the wave
		 \task \textbf{The operating frequency}
		 \task Final use of the signal
		 \end{tasks} 
	 \item \emph{(\protect\tikz[baseline=-0.5ex]{\protect\draw[pink, very thick] (0,0) -- (0.5,0);})} Divergence ($D$) is the reduction of the reflection coefficient due to the effect of \_\_\_\_ of the Earth:
	 \begin{tasks}(4)
		 \task Divergence
		 \task Concavity
		 \task Convergence
		 \task \textbf{Convexity}
		 \end{tasks} 
	 \item \emph{(\protect\tikz[baseline=-0.5ex]{\protect\draw[pink, very thick] (0,0) -- (0.5,0);})} Ionospheric propagation has (indicate the false option):
	 \begin{tasks}(4)
		 \task Range of useful values in $f$ (MUF, LUF)
		 \task \textbf{Low interferences}
		 \task Random nature due to ionosphere
		 \task Very high natural noise
		 \end{tasks} 
	 \item \emph{(\protect\tikz[baseline=-0.5ex]{\protect\draw[pink, very thick] (0,0) -- (0.5,0);})} In diffraction-based propagation, the signal remains essentially unaffected provided that the obstacle obstructs no more than \_\_\_\_\% of the radius of the first Fresnel zone: 
	 \begin{tasks}(4)
		 \task 70
		 \task \textbf{40}
		 \task 70
		 \task 100
		 \end{tasks} 
	 \item \emph{(\protect\tikz[baseline=-0.5ex]{\protect\draw[pink, very thick] (0,0) -- (0.5,0);})} The ionospheric wave operates at frequencies:
	 \begin{tasks}(4)
		 \task LF
		 \task \textbf{ MF}
		 \task VHF
		 \task UHF
		 \end{tasks} 
	 \item \emph{(\protect\tikz[baseline=-0.5ex]{\protect\draw[pink, very thick] (0,0) -- (0.5,0);})} Lex losses are the additional losses in signal power that occur in a real propagation environment compared to the ideal free-space path loss (FSPL). What is the name for this ``Lex" abbreviation?: 
	 \begin{tasks}(4)
		 \task Extraordinary losses
		 \task \textbf{Excess losses}
		 \task Extra losses
		 \task X-men losses
		 \end{tasks} 
	 \item \emph{(\protect\tikz[baseline=-0.5ex]{\protect\draw[pink, very thick] (0,0) -- (0.5,0);})} The surface wave model is valid up to:
	 \begin{tasks}(4)
		 \task 1 GHz
		 \task 50 MHz
		 \task \textbf{150 MHz}
		 \task 100 kHz
		 \end{tasks} 
	 \item \emph{(\protect\tikz[baseline=-0.5ex]{\protect\draw[pink, very thick] (0,0) -- (0.5,0);})} The angle at which the parallel polarization component to the plane is canceled is called:
	 \begin{tasks}(4)
		 \task Polarized angle
		 \task \textbf{Brewster angle}
		 \task Depolarized angle
		 \task Coldbrew angle
		 \end{tasks} 
	 \item \emph{(\protect\tikz[baseline=-0.5ex]{\protect\draw[pink, very thick] (0,0) -- (0.5,0);})} In Fresnel diffraction, the critical point, if there are no obstacles in the path, is:
	 \begin{tasks}(4)
		 \task \textbf{The center ($d/2$)}
		 \task One quarter ($d/4$)
		 \task Three quarters ($3d/4$)
		 \task Radio horizon
		 \end{tasks} 
	 \item \emph{(\protect\tikz[baseline=-0.5ex]{\protect\draw[pink, very thick] (0,0) -- (0.5,0);})} Fading of “power” or factor $k$ is controlled by:
	 \begin{tasks}(4)
		 \task By increasing the radiated power
		 \task \textbf{By modifying antenna heights}
		 \task Avoiding natural attenuations
		 \task By modifying antenna polarization
		 \end{tasks} 
	 \item \emph{(\protect\tikz[baseline=-0.5ex]{\protect\draw[pink, very thick] (0,0) -- (0.5,0);})} The symbol $\gamma$ is called \_\_\_\_ and is the criterion of whom?: 
	 \begin{tasks}(4)
		 \task Pasta, Rayleigh criterion
		 \task Pasta, roughness criterion
		 \task Gamma, roughness criterion
		 \task \textbf{Gamma, Rayleigh criterion}
		 \end{tasks} 
	 \item \emph{(\protect\tikz[baseline=-0.5ex]{\protect\draw[pink, very thick] (0,0) -- (0.5,0);})} We can apply the flat Earth model because: 
	 \begin{tasks}(4)
		 \task The Earth is indeed flat
		 \task At low frequencies, Earth curvature can be neglected
		 \task \textbf{Earth curvature can be neglected at short distances}
		 \task Only if $d \approx h_t,h_r$
		 \end{tasks} 
	 \item \emph{(\protect\tikz[baseline=-0.5ex]{\protect\draw[pink, very thick] (0,0) -- (0.5,0);})} The received powers at the receivers are modeled with:
	 \begin{tasks}(4)
		 \task Maxwell’s equations
		 \task Circuits
		 \task \textbf{Statistical distribution}
		 \task Artificial Intelligence
		 \end{tasks} 
	 \item \emph{(\protect\tikz[baseline=-0.5ex]{\protect\draw[pink, very thick] (0,0) -- (0.5,0);})} In the curved Earth model:
	 \begin{tasks}(4)
		 \task The distance can be greater than the radio horizon
		 \task We work with 2 curves
		 \task \textbf{Earth curvature is appreciable}
		 \task Equivalent to the flat Earth model
		 \end{tasks} 
	 \item \emph{(\protect\tikz[baseline=-0.5ex]{\protect\draw[pink, very thick] (0,0) -- (0.5,0);})} The basic losses in flat Earth and curved Earth are: (TC = Basic transmission loss in the curved Earth model; TP = Basic transmission loss in the flat Earth model)
	 \begin{tasks}(4)
		 \task The formula is TC = 2 TP
		 \task \textbf{They are the same}
		 \task The formula is 2TC = TP
		 \task The formula is (1/$\sqrt{2}$) TC = TP
		 \end{tasks} 
	 \item \emph{(\protect\tikz[baseline=-0.5ex]{\protect\draw[pink, very thick] (0,0) -- (0.5,0);})} P.1546, Okumura-Hata, COST 231 are:
	 \begin{tasks}(4)
		 \task Point-to-point propagation models
		 \task Practical propagation prediction models
		 \task Exact empirical propagation models
		 \task \textbf{Empirical propagation prediction methods}
		 \end{tasks} 
	 \item \emph{(\protect\tikz[baseline=-0.5ex]{\protect\draw[pink, very thick] (0,0) -- (0.5,0);})} The basic free-space radiation losses do NOT depend on:
	 \begin{tasks}(4)
		 \task The operating frequency
		 \task Obstacles in the path
		 \task \textbf{Transmitting antenna gain}
		 \task Wave polarization
		 \end{tasks} 
	 \item \emph{(\protect\tikz[baseline=-0.5ex]{\protect\draw[pink, very thick] (0,0) -- (0.5,0);})} The tropospheric wave operates at frequencies:
	 \begin{tasks}(4)
		 \task LF
		 \task MF
		 \task \textbf{VHF}
		 \task HF
		 \end{tasks} 
	 \item \emph{(\protect\tikz[baseline=-0.5ex]{\protect\draw[pink, very thick] (0,0) -- (0.5,0);})} If for an isotropic antenna it is EIRP, for an SVA it is:
	 \begin{tasks}(4)
		 \task ERMP
		 \task \textbf{EMRP}
		 \task PRME
		 \task PREM
		 \end{tasks} 
	 \item \emph{(\protect\tikz[baseline=-0.5ex]{\protect\draw[pink, very thick] (0,0) -- (0.5,0);})} In multipath fading, if no dominant component exists, a statistical distribution type \_\_\_\_ is used:
	 \begin{tasks}(4)
		 \task Rice
		 \task Binomial
		 \task Normal
		 \task \textbf{Rayleigh}
		 \end{tasks} 
	 \item \emph{(\protect\tikz[baseline=-0.5ex]{\protect\draw[pink, very thick] (0,0) -- (0.5,0);})} Pearson’s depression deals with the decrease of:
	 \begin{tasks}(4)
		 \task Field values
		 \task Attenuation
		 \task \textbf{Received power}
		 \task Money
		 \end{tasks} 
	 \item \emph{(\protect\tikz[baseline=-0.5ex]{\protect\draw[pink, very thick] (0,0) -- (0.5,0);})} Due to small changes in the refractive index ($n$), it is preferable to work with: 
	 \begin{tasks}(4)
		 \task Intrinsic impedance of free space
		 \task Capacitance
		 \task Earth curvature
		 \task \textbf{Co-index ($N$)}
		 \end{tasks} 
	 \item \emph{(\protect\tikz[baseline=-0.5ex]{\protect\draw[pink, very thick] (0,0) -- (0.5,0);})} Fading is any decrease in power relative to the nominal value:
	 \begin{tasks}(4)
		 \task \textbf{Yes}
		 \task No
		 \task It is only partial decrease
		 \task What faded were her feelings
		 \end{tasks} 
	 \item \emph{(\protect\tikz[baseline=-0.5ex]{\protect\draw[pink, very thick] (0,0) -- (0.5,0);})} If the effective Earth radius factor ($k$) is greater than 1, it is called:
	 \begin{tasks}(4)
		 \task \textbf{Superrefractive}
		 \task Subrefractive
		 \task Subrefractive
		 \task Ultrarefractive
		 \end{tasks} 
	 \item \emph{(\protect\tikz[baseline=-0.5ex]{\protect\draw[pink, very thick] (0,0) -- (0.5,0);})} In curved Earth, the reflection coefficient ($R$) is modified with:
	 \begin{tasks}(4)
		 \task \textbf{Divergence and roughness}
		 \task Divergence and path difference
		 \task Propagation losses
		 \task Flat Earth model parameters
		 \end{tasks} 
	 \item \emph{(\protect\tikz[baseline=-0.5ex]{\protect\draw[pink, very thick] (0,0) -- (0.5,0);})} Tropospheric propagation is used when:
	 \begin{tasks}(4)
		 \task \textbf{We can omit the obstacle}
		 \task At frequencies $>$ 150 MHz
		 \task Antenna length $\approx \lambda$
		 \task Receiving antenna is beyond radio horizon
		 \end{tasks} 
	 \item \emph{(\protect\tikz[baseline=-0.5ex]{\protect\draw[pink, very thick] (0,0) -- (0.5,0);})} If the effective Earth radius factor ($k$) is less than 1, it is called:
	 \begin{tasks}(4)
		 \task Superrefractive
		 \task \textbf{Subrefractive}
		 \task Subrefractive
		 \task Ultrarefractive
		 \end{tasks} 
	 \item \emph{(\protect\tikz[baseline=-0.5ex]{\protect\draw[pink, very thick] (0,0) -- (0.5,0);})} The reflection coefficient ($R$) does not depend on:
	 \begin{tasks}(4)
		 \task The angle of incidence
		 \task Polarization
		 \task Electrical characteristics of the ground
		 \task \textbf{Radiated power}
		 \end{tasks} 
	 \item \emph{(\protect\tikz[baseline=-0.5ex]{\protect\draw[pink, very thick] (0,0) -- (0.5,0);})} If the effective Earth radius factor ($k$) is 4/3, it is called:
	 \begin{tasks}(4)
		 \task Isotropic
		 \task \textbf{Standard}
		 \task Basic
		 \task Normal
		 \end{tasks} 
	 \item \emph{(\protect\tikz[baseline=-0.5ex]{\protect\draw[pink, very thick] (0,0) -- (0.5,0);})} Indicate the polarization of the antenna used in Surface Wave propagation:
	 \begin{tasks}(4)
		 \task \textbf{Vertical linear polarization}
		 \task Horizontal linear polarization
		 \task Left-hand circular polarization
		 \task Right-hand polarization
		 \end{tasks} 
	 \item \emph{(\protect\tikz[baseline=-0.5ex]{\protect\draw[pink, very thick] (0,0) -- (0.5,0);})} The ITU-R diffraction graphs are known as $L_D$; they are used for what type of obstacles?:
	 \begin{tasks}(4)
		 \task \textbf{Sharp obstacle}
		 \task Rounded obstacle
		 \task Not used for obstacle
		 \task Multiple obstacle
		 \end{tasks} 
	 \item \emph{(\protect\tikz[baseline=-0.5ex]{\protect\draw[pink, very thick] (0,0) -- (0.5,0);})} In surface wave, a radio wave travels: 
	 \begin{tasks}(4)
		 \task \textbf{Along the Earth's surface, following its curvature and being influenced by its conductivity}
		 \task Along the Earth's surface, following the clouds shape
		 \task Along the Earth's the ionospheric layer
		 \task Waves do not exist
		 \end{tasks} 
	 \item \emph{(\protect\tikz[baseline=-0.5ex]{\protect\draw[pink, very thick] (0,0) -- (0.5,0);})} In the ionospheric wave, the wave:
	 \begin{tasks}(4)
		 \task Passes through the ionosphere
		 \task Refracts in the ionosphere and returns to origin
		 \task \textbf{Refracts in the ionosphere and returns to ground}
		 \task It never comes back, like her
		 \end{tasks} 
	 \item \emph{(\protect\tikz[baseline=-0.5ex]{\protect\draw[pink, very thick] (0,0) -- (0.5,0);})} Since solving the real problem is very complex, as an alternative we use:
	 \begin{tasks}(4)
		 \task Maxwell’s equations
		 \task Mathematical tools like vector potential A and scalar V
		 \task \textbf{Simplified models based on geometric optics}
		 \task All of the above
		 \end{tasks} 
	 \item \emph{(\protect\tikz[baseline=-0.5ex]{\protect\draw[pink, very thick] (0,0) -- (0.5,0);})} The relative permittivity $\varepsilon_r$ and conductivity $\sigma$ have graphs with curves that represent:
	 \begin{tasks}(4)
		 \task The frequency
		 \task Antenna height
		 \task Polarization
		 \task \textbf{Soil type}
		 \end{tasks} 
	 \item \emph{(\protect\tikz[baseline=-0.5ex]{\protect\draw[pink, very thick] (0,0) -- (0.5,0);})} In the flat Earth model, the basic losses increase with the fourth power of the distance:
	 \begin{tasks}(4)
		 \task Always
		 \task Only at 150 MHz
		 \task \textbf{In the working area}
		 \task Above 150 MHz
		 \end{tasks} 
	 \item \emph{(\protect\tikz[baseline=-0.5ex]{\protect\draw[pink, very thick] (0,0) -- (0.5,0);})} In the fundamental propagation equation written in its compact form: $e = e_{0} \left[ 1 + R \, e^{-j\Delta} + (1 - R) \, A \, e^{-j\Delta} \right]$, the 3 ray components in order are: (Direct ray = DR; Reflected ray = RR; Surface wave = SW): 
	 \begin{tasks}(4)
		 \task \textbf{DR, RR, SW}
		 \task DR, RR, Diffraction
		 \task Fading, DR, SW
		 \task DR, SW, RR
		 \end{tasks} 
	 \item \emph{(\protect\tikz[baseline=-0.5ex]{\protect\draw[pink, very thick] (0,0) -- (0.5,0);})} If the effective Earth radius factor ($k$) is less than 1:
	 \begin{tasks}(4)
		 \task $R_{(eff.)} > R_{(real)}$
		 \task \bm{$R_{(eff.)} < R_{(real)}$}
		 \task $R_{(eff.)} = R_{(real)}$
		 \task $R_{(eff.)} = \sqrt{R_{(real)}}$
		 \end{tasks} 
	 \item \emph{(\protect\tikz[baseline=-0.5ex]{\protect\draw[pink, very thick] (0,0) -- (0.5,0);})} Lex losses (Excess losses) follow the relation: 
	 \begin{tasks}(4)
		 \task $e_0 / |e|$
		 \task $|e| / e_0$
		 \task \bm{$\Big||e| / e_0\Big|^2$}
		 \task $\Big|\epsilon_0 / |\epsilon|\Big|^2$
		 \end{tasks} 
	 \item \emph{(\protect\tikz[baseline=-0.5ex]{\protect\draw[pink, very thick] (0,0) -- (0.5,0);})} The radio horizon distance is: 
	 \begin{tasks}(4)
		 \task $3.57 \sqrt{k \cdot h_t}$ [m]
		 \task $3.57 \sqrt{k \cdot h_r}$ [m]
		 \task \bm{$3.57 \sqrt{k} \Big[\sqrt{h_t} + \sqrt{h_r}\Big]$} \textbf{[km]}
		 \task $3.57 \sqrt{k_{1,2} \cdot h_{t,r}}$ [km]
		 \end{tasks} 
	 \item \emph{(\protect\tikz[baseline=-0.5ex]{\protect\draw[yellow2, very thick] (0,0) -- (0.5,0);})} A current flowing vertically in the positive direction generates a magnetic field in the direction:
	 \begin{tasks}(4)
		 \task Clockwise
		 \task \textbf{Counterclockwise}
		 \task Vertical
		 \task Horizontal
		 \end{tasks} 
	 \item \emph{(\protect\tikz[baseline=-0.5ex]{\protect\draw[yellow2, very thick] (0,0) -- (0.5,0);})} The current density $J$ has units of:
	 \begin{tasks}(4)
		 \task A/m
		 \task A
		 \task \textbf{A/m$^2$}
		 \task A/m$^3$
		 \end{tasks} 
	 \item \emph{(\protect\tikz[baseline=-0.5ex]{\protect\draw[yellow2, very thick] (0,0) -- (0.5,0);})} The displacement current is a:
	 \begin{tasks}(4)
		 \task Conduction current
		 \task Proton flux
		 \task \textbf{Electric field variation}
		 \task Magnetic field variation
		 \end{tasks} 
	 \item \emph{(\protect\tikz[baseline=-0.5ex]{\protect\draw[yellow2, very thick] (0,0) -- (0.5,0);})} The refractive index is defined as the ratio (in this order) between:
	 \begin{tasks}(4)
		 \task \textbf{Speed of light in vacuum / speed in the medium}
		 \task Medium speed / speed of light in vacuum
		 \task Electric field / Magnetic field
		 \task Magnetic field / Electric field
		 \end{tasks} 
	 \item \emph{(\protect\tikz[baseline=-0.5ex]{\protect\draw[yellow2, very thick] (0,0) -- (0.5,0);})} Inductance has units of:
	 \begin{tasks}(4)
		 \task Farad
		 \task Ampere $\times$ second
		 \task \textbf{Weber / Ampere}
		 \task Farad / Ohm
		 \end{tasks} 
	 \item \emph{(\protect\tikz[baseline=-0.5ex]{\protect\draw[yellow2, very thick] (0,0) -- (0.5,0);})} A short-circuited transmission line has an input impedance:
	 \begin{tasks}(4)
		 \task Real
		 \task \textbf{Imaginary}
		 \task Complex
		 \task Integer
		 \end{tasks} 
	 \item \emph{(\protect\tikz[baseline=-0.5ex]{\protect\draw[yellow2, very thick] (0,0) -- (0.5,0);})} The following type of particle experiences a force and moves in the direction of the electric field:
	 \begin{tasks}(4)
		 \task \textbf{Proton}
		 \task Electron
		 \task Neutron
		 \task All of the above
		 \end{tasks} 
	 \item \emph{(\protect\tikz[baseline=-0.5ex]{\protect\draw[yellow2, very thick] (0,0) -- (0.5,0);})} The admittance of a wave relates (in this order):
	 \begin{tasks}(4)
		 \task $H$ / $E$
		 \task \textbf{\textit{E}} / \textbf{\textit{H}}
		 \task $E \times H$
		 \task Other
		 \end{tasks} 
	 \item \emph{(\protect\tikz[baseline=-0.5ex]{\protect\draw[yellow2, very thick] (0,0) -- (0.5,0);})} In electrostatics, the electric field and potential V are related as:
	 \begin{tasks}(4)
		 \task $\vec{E} = -\vec{B} + \nabla V$
		 \task \bm{$\vec{E} = -\nabla V$}
		 \task $\vec{E} = -\int V \cdot dl$
		 \task $\vec{E} = -V$
		 \end{tasks} 
	 \item \emph{(\protect\tikz[baseline=-0.5ex]{\protect\draw[yellow2, very thick] (0,0) -- (0.5,0);})} Magnetic permeability is denoted by the letter:
	 \begin{tasks}(4)
		 \task $\varepsilon$
		 \task \bm{$\mu$}
		 \task $\eta$
		 \task $\Gamma$
		 \end{tasks} 
	 \item \emph{(\protect\tikz[baseline=-0.5ex]{\protect\draw[yellow2, very thick] (0,0) -- (0.5,0);})} Stokes’ theorem states that:
	 \begin{tasks}(4)
		 \task \bm{$\int\!\!\int_S (\nabla \times \vec{E}) \cdot d\vec{S}$} \\ \bm{$= \oint_C \vec{E} \cdot d\vec{l}$}
		 \task $\int\!\!\int\!\!\int_V (\nabla \times \vec{E}) \cdot d\vec{S} \\= \oint \!\!\oint_S \vec{E} \cdot d\vec{l}$
		 \task $\int\!\!\int_S (\nabla \cdot \vec{E})  dS \\= \oint_C \vec{E} \cdot d\vec{l}$
		 \task Other
		 \end{tasks} 
	 \item \emph{(\protect\tikz[baseline=-0.5ex]{\protect\draw[yellow2, very thick] (0,0) -- (0.5,0);})} A solenoidal field has associated a: 
	 \begin{tasks}(4)
		 \task \textbf{Null divergence}
		 \task Null curl
		 \task Null Laplacian
		 \task Other
		 \end{tasks} 
	 \item \emph{(\protect\tikz[baseline=-0.5ex]{\protect\draw[yellow2, very thick] (0,0) -- (0.5,0);})} Gauss’s law for the electric field is: 
	 \begin{tasks}(4)
		 \task $\nabla \cdot \vec{E} = \rho/\varepsilon$
		 \task $\int \int \vec{E} \cdot d\vec{S} = Q/\varepsilon$
		 \task \textbf{The two previous ones}
		 \task $\nabla \cdot \vec{B} = 0$
		 \end{tasks} 
	 \item \emph{(\protect\tikz[baseline=-0.5ex]{\protect\draw[yellow2, very thick] (0,0) -- (0.5,0);})} Which of the following equations is NOT one of Maxwell’s equations?:
	 \begin{tasks}(4)
		 \task Gauss’s theorem
		 \task \textbf{Continuity equation}
		 \task Ampère–Maxwell's law
		 \task Faraday–Lenz's law
		 \end{tasks} 
	 \item \emph{(\protect\tikz[baseline=-0.5ex]{\protect\draw[yellow2, very thick] (0,0) -- (0.5,0);})} The following vector operation does NOT appear directly in Maxwell’s equations:
	 \begin{tasks}(4)
		 \task Divergence
		 \task Curl
		 \task \textbf{Gradient}
		 \task Partial derivative (time)
		 \end{tasks} 
	 \item \emph{(\protect\tikz[baseline=-0.5ex]{\protect\draw[yellow2, very thick] (0,0) -- (0.5,0);})} According to the physical meaning of divergence, if it is greater than 0, this means: 
	 \begin{tasks}(4)
		 \task Null of electric field
		 \task \textbf{Source of electric field}
		 \task Sink of electric field
		 \task Home of electric field
		 \end{tasks} 
	 \item \emph{(\protect\tikz[baseline=-0.5ex]{\protect\draw[yellow2, very thick] (0,0) -- (0.5,0);})} Two charges $q_1$ and $q_2$ separated by a distance $r$ experience a force proportional to:
	 \begin{tasks}(4)
		 \task $1/r$
		 \task \bm{$1/r^2$}
		 \task $1/r^3$
		 \task $1/r + 1/r^2 + 1/r^3$
		 \end{tasks} 
	 \item \emph{(\protect\tikz[baseline=-0.5ex]{\protect\draw[yellow2, very thick] (0,0) -- (0.5,0);})} Two currents flowing in the same direction generate a force:
	 \begin{tasks}(4)
		 \task \textbf{Attractive}
		 \task Repulsive
		 \task Depends
		 \task On what does it depend?
		 \end{tasks} 
	 \item \emph{(\protect\tikz[baseline=-0.5ex]{\protect\draw[yellow2, very thick] (0,0) -- (0.5,0);})} An alternative law to Ampère’s law for studying magnetic fields is:
	 \begin{tasks}(4)
		 \task Coulomb’s law
		 \task Gauss’s law
		 \task \textbf{Biot–Savart law}
		 \task Law of the strongest
		 \end{tasks} 
	 \item \emph{(\protect\tikz[baseline=-0.5ex]{\protect\draw[yellow2, very thick] (0,0) -- (0.5,0);})} Gauss’s law for the magnetic field tells us that:
	 \begin{tasks}(4)
		 \task \textbf{Magnetic monopoles do not exist}
		 \task Magnetic monopoles exist
		 \task Electric monopoles do not exist
		 \task Her love does not exist
		 \end{tasks} 
	 \item \emph{(\protect\tikz[baseline=-0.5ex]{\protect\draw[yellow2, very thick] (0,0) -- (0.5,0);})} Ohm’s law can also be expressed as:
	 \begin{tasks}(4)
		 \task $I = V · R$
		 \task $\boldsymbol{\vec{J} = \sigma \vec{E}}$
		 \task $\vec{E} = \sigma \vec{J}$
		 \task $\vec{J} = \sigma \vec{I}$
		 \end{tasks} 
	 \item \emph{(\protect\tikz[baseline=-0.5ex]{\protect\draw[yellow2, very thick] (0,0) -- (0.5,0);})} Two currents flowing in opposite directions generate a force:
	 \begin{tasks}(4)
		 \task Attractive
		 \task \textbf{Repulsive}
		 \task Depends on the day
		 \task Depends on atmospheric conditions
		 \end{tasks} 
	 \item \emph{(\protect\tikz[baseline=-0.5ex]{\protect\draw[yellow2, very thick] (0,0) -- (0.5,0);})} Voltage can be calculated as:
	 \begin{tasks}(4)
		 \task Derivative of $E$-field
		 \task Surface integral of $E$-field
		 \task \textbf{Line integral of $E$-field}
		 \task Volume integral of $E$-field
		 \end{tasks} 
	 \item \emph{(\protect\tikz[baseline=-0.5ex]{\protect\draw[yellow2, very thick] (0,0) -- (0.5,0);})} The conductivity of a perfect electric conductor metal is:
	 \begin{tasks}(4)
		 \task $\sigma = 0$
		 \task $\sigma = 8.85 \cdot 10^{-12}$
		 \task \textbf{Infinity}
		 \task $\sigma = 5.7 \cdot 10^{7}$
		 \end{tasks} 
	 \item \emph{(\protect\tikz[baseline=-0.5ex]{\protect\draw[yellow2, very thick] (0,0) -- (0.5,0);})} In the right-hand rule applied to magnetic forces, the index and middle fingers represent, respectively:
	 \begin{tasks}(4)
		 \task \textbf{Velocity and magnetic field}
		 \task Magnetic field and velocity
		 \task Force and velocity
		 \task Velocity and force
		 \end{tasks} 
	 \item \emph{(\protect\tikz[baseline=-0.5ex]{\protect\draw[yellow2, very thick] (0,0) -- (0.5,0);})} The magnetic field $\vec{B}$ is calculated as:
	 \begin{tasks}(4)
		 \task Curl of the electric field
		 \task \textbf{Curl of magnetic vector potential}
		 \task Curl of voltage
		 \task Curl of charge
		 \end{tasks} 
	 \item \emph{(\protect\tikz[baseline=-0.5ex]{\protect\draw[yellow2, very thick] (0,0) -- (0.5,0);})} Hydroelectric power plants transform motion into electrical energy thanks to the principle of:
	 \begin{tasks}(4)
		 \task \textbf{Induction}
		 \task Contraction
		 \task Reduction
		 \task Cardboard
		 \end{tasks} 
	 \item \emph{(\protect\tikz[baseline=-0.5ex]{\protect\draw[yellow2, very thick] (0,0) -- (0.5,0);})} The divergence theorem is applied to convert the following equation from differential to integral form:
	 \begin{tasks}(4)
		 \task Ampère's
		 \task Faraday's
		 \task Coulomb's
		 \task \textbf{Gauss's}
		 \end{tasks} 
	 \item \emph{(\protect\tikz[baseline=-0.5ex]{\protect\draw[yellow2, very thick] (0,0) -- (0.5,0);})} In a linear, isotropic and non-dispersive medium, the fields $\vec{D}$ and $\vec{E}$ are related as:
	 \begin{tasks}(4)
		 \task $\vec{E} = v  \vec{D}$
		 \task \bm{$\vec{D} = \varepsilon \vec{E}$}
		 \task $\vec{E} = \varepsilon \vec{D}$
		 \task $\vec{D} = \varepsilon (\vec{E} + \vec{B})$
		 \end{tasks} 
	 \item \emph{(\protect\tikz[baseline=-0.5ex]{\protect\draw[yellow2, very thick] (0,0) -- (0.5,0);})} The fundamental mode in a transmission line is:
	 \begin{tasks}(4)
		 \task TE
		 \task TM
		 \task \textbf{TEM}
		 \task TE10
		 \end{tasks} 
	 \item \emph{(\protect\tikz[baseline=-0.5ex]{\protect\draw[yellow2, very thick] (0,0) -- (0.5,0);})} A conservative field is associated with the operation:
	 \begin{tasks}(4)
		 \task $\nabla \cdot \vec{J} = 0$
		 \task \bm{$\nabla \times \vec{J} = 0$}
		 \task $\nabla^2 \vec{J} = \vec{0}$
		 \task Other
		 \end{tasks} 
	 \item \emph{(\protect\tikz[baseline=-0.5ex]{\protect\draw[yellow2, very thick] (0,0) -- (0.5,0);})} The capacitances of two capacitors in parallel are:
	 \begin{tasks}(4)
		 \task Multiplied
		 \task \textbf{Added}
		 \task Divided
		 \task Operated as two resistances in parallel
		 \end{tasks} 
	 \item \emph{(\protect\tikz[baseline=-0.5ex]{\protect\draw[yellow2, very thick] (0,0) -- (0.5,0);})} The wave number is calculated as:
	 \begin{tasks}(4)
		 \task $k = \omega / v$
		 \task $k = 2\pi / \lambda$
		 \task $k = 2\pi f / v$
		 \task \textbf{All of the above}
		 \end{tasks} 
	 \item \emph{(\protect\tikz[baseline=-0.5ex]{\protect\draw[yellow2, very thick] (0,0) -- (0.5,0);})} The displacement current generated by a constant electric field is:
	 \begin{tasks}(4)
		 \task Constant
		 \task \textbf{Zero}
		 \task Its derivative
		 \task Its integral
		 \end{tasks} 
	 \item \emph{(\protect\tikz[baseline=-0.5ex]{\protect\draw[yellow2, very thick] (0,0) -- (0.5,0);})} One Coulomb contains the following number of electrons:
	 \begin{tasks}(4)
		 \task $2$
		 \task $300,000,000$
		 \task $9 \cdot 10^{16}$
		 \task \bm{$6.24 \cdot 10^{18}$}
		 \end{tasks} 
	 \item \emph{(\protect\tikz[baseline=-0.5ex]{\protect\draw[yellow2, very thick] (0,0) -- (0.5,0);})} In a linear, isotropic and non-dispersive medium, the fields B and H are related as:
	 \begin{tasks}(4)
		 \task \bm{$\vec{B} = \mu \vec{H}$}
		 \task $\vec{H} = \mu \vec{B}$
		 \task $\vec{H} = \varepsilon \mu \vec{B}$
		 \task $\vec{B} = \mu \varepsilon \vec{H}$
		 \end{tasks} 
	 \item \emph{(\protect\tikz[baseline=-0.5ex]{\protect\draw[yellow2, very thick] (0,0) -- (0.5,0);})} The impedance of free space is (Ohms):
	 \begin{tasks}(4)
		 \task $0$
		 \task $33$
		 \task $119\pi$
		 \task \bm{$120\pi$}
		 \end{tasks} 
	 \item \emph{(\protect\tikz[baseline=-0.5ex]{\protect\draw[yellow2, very thick] (0,0) -- (0.5,0);})} In spherical coordinates:
	 \begin{tasks}(4)
		 \task $\hat{r} \times \hat{\phi} = \hat{\theta}$
		 \task \bm{$\hat{r} \times \hat{\theta} = \hat{\phi}$}
		 \task $\hat{r} \times \hat{r} = \hat{r}$
		 \task Other
		 \end{tasks} 
	 \item \emph{(\protect\tikz[baseline=-0.5ex]{\protect\draw[yellow2, very thick] (0,0) -- (0.5,0);})} The divergence theorem is also known as:
	 \begin{tasks}(4)
		 \task Jacobi's theorem
		 \task Bernoulli's theorem
		 \task Sanmartín's theorem
		 \task \textbf{Gauss's theorem}
		 \end{tasks} 
	 \item \emph{(\protect\tikz[baseline=-0.5ex]{\protect\draw[yellow2, very thick] (0,0) -- (0.5,0);})} The electric field has units of:
	 \begin{tasks}(4)
		 \task Meter / volt
		 \task Volt $\times$ meter
		 \task Newton $\times$ Coulomb
		 \task \textbf{Newton / Coulomb}
		 \end{tasks} 
	 \item \emph{(\protect\tikz[baseline=-0.5ex]{\protect\draw[yellow2, very thick] (0,0) -- (0.5,0);})} The Farad is defined as:
	 \begin{tasks}(4)
		 \task coulomb × volt
		 \task Coulomb / ampere
		 \task \textbf{Coulomb / volt}
		 \task Volt / coulomb
		 \end{tasks} 
	 \item \emph{(\protect\tikz[baseline=-0.5ex]{\protect\draw[yellow2, very thick] (0,0) -- (0.5,0);})} Faraday’s law is:
	 \begin{tasks}(4)
		 \task $\nabla \cdot \vec{E} = \rho/\varepsilon$
		 \task $\nabla \cdot \vec{B} = 0$
		 \task $\nabla \times \vec{B} = \mu \vec{J} + \mu \varepsilon (\partial \vec{E}/\partial t)$
		 \task \bm{$\nabla \times \vec{E} = - \partial \vec{B}/\partial t$}
		 \end{tasks} 
	 \item \emph{(\protect\tikz[baseline=-0.5ex]{\protect\draw[yellow2, very thick] (0,0) -- (0.5,0);})} The area of a sphere is:
	 \begin{tasks}(4)
		 \task $2\pi r^2$
		 \task $3\pi r^2$
		 \task \bm{$4\pi r^2$}
		 \task $5\pi r^2$
		 \end{tasks} 
	 \item \emph{(\protect\tikz[baseline=-0.5ex]{\protect\draw[yellow2, very thick] (0,0) -- (0.5,0);})} The unit of capacitance is:
	 \begin{tasks}(4)
		 \task Henry
		 \task Ohm
		 \task \textbf{Farad}
		 \task Tungsten
		 \end{tasks} 
	 \item \emph{(\protect\tikz[baseline=-0.5ex]{\protect\draw[yellow2, very thick] (0,0) -- (0.5,0);})} The continuity equation of charge-current states that:
	 \begin{tasks}(4)
		 \task $\nabla \cdot \vec{J} = k$
		 \task $\nabla \times \vec{J} = -\partial \rho / \partial t$
		 \task \bm{$\nabla \cdot \vec{J} = -\partial \rho/\partial t$}
		 \task $\nabla \times \vec{J} = -\nabla \cdot \vec{J}$
		 \end{tasks} 
	 \item \emph{(\protect\tikz[baseline=-0.5ex]{\protect\draw[yellow2, very thick] (0,0) -- (0.5,0);})} A magnetic flux is canceled when vector B and the differential surface vector form:
	 \begin{tasks}(4)
		 \task \textbf{90°}
		 \task 0°
		 \task 45°
		 \task None of the above
		 \end{tasks} 
	 \item \emph{(\protect\tikz[baseline=-0.5ex]{\protect\draw[yellow2, very thick] (0,0) -- (0.5,0);})} Which of the following metals has magnetic properties?
	 \begin{tasks}(4)
		 \task Aluminum
		 \task Copper
		 \task \textbf{Iron}
		 \task Silver
		 \end{tasks} 
	 \item \emph{(\protect\tikz[baseline=-0.5ex]{\protect\draw[blue, very thick] (0,0) -- (0.5,0);})} Indicate which is NOT a satellite:
	 \begin{tasks}(4)
		 \task Hispasat
		 \task Astra
		 \task \textbf{Euskadi}
		 \task Eutelsat
		 \end{tasks} 
	 \item \emph{(\protect\tikz[baseline=-0.5ex]{\protect\draw[blue, very thick] (0,0) -- (0.5,0);})} The inventors of the Yagi-Uda antenna were:
	 \begin{tasks}(4)
		 \task Chinese
		 \task Spaniards
		 \task Thai
		 \task \textbf{Japanese}
		 \end{tasks} 
	 \item \emph{(\protect\tikz[baseline=-0.5ex]{\protect\draw[blue, very thick] (0,0) -- (0.5,0);})} Which metal has the greatest magnetic permeability?
	 \begin{tasks}(4)
		 \task Nickel
		 \task Cobalt
		 \task Aluminum
		 \task \textbf{Iron}
		 \end{tasks} 
	 \item \emph{(\protect\tikz[baseline=-0.5ex]{\protect\draw[blue, very thick] (0,0) -- (0.5,0);})} Which of the following antennas is an electrically short monopole antenna?
	 \begin{tasks}(4)
		 \task Patch antenna
		 \task Linear horn antenna
		 \task \textbf{Rubber duck antenna}
		 \task Vivaldi antenna
		 \end{tasks} 
	 \item \emph{(\protect\tikz[baseline=-0.5ex]{\protect\draw[blue, very thick] (0,0) -- (0.5,0);})} The nabla operator is also known by the name:
	 \begin{tasks}(4)
		 \task Aleph
		 \task Alpha
		 \task Atlas
		 \task \textbf{Atled}
		 \end{tasks} 
	 \item \emph{(\protect\tikz[baseline=-0.5ex]{\protect\draw[blue, very thick] (0,0) -- (0.5,0);})} In the 2021 videogame “It Takes Two,” in the Lunar Baboon chapter, which antenna appears?
	 \begin{tasks}(4)
		 \task Satellite
		 \task Corrugated horn
		 \task \textbf{Parabolic reflector}
		 \task Monopole array
		 \end{tasks} 
	 \item \emph{(\protect\tikz[baseline=-0.5ex]{\protect\draw[blue, very thick] (0,0) -- (0.5,0);})} The “rubber duck” antenna gets its name from a comment by:
	 \begin{tasks}(4)
		 \task President John F. Kennedy
		 \task \textbf{President J.F. Kennedy’s daughter}
		 \task J.F. Kennedy’s secret agent
		 \task Such an antenna does not exist
		 \end{tasks} 
	 \item \emph{(\protect\tikz[baseline=-0.5ex]{\protect\draw[blue, very thick] (0,0) -- (0.5,0);})} The creator of the Vivaldi antenna named it so because:
	 \begin{tasks}(4)
		 \task His girlfriend’s surname was that
		 \task His mother’s surname was that
		 \task \textbf{He liked classical music}
		 \task It actually was Antonio Vivaldi
		 \end{tasks} 
	 \item \emph{(\protect\tikz[baseline=-0.5ex]{\protect\draw[blue, very thick] (0,0) -- (0.5,0);})} The scientist responsible for reducing Maxwell’s equations from 20 to only 4 was:
	 \begin{tasks}(4)
		 \task Faraday
		 \task Planck
		 \task \textbf{Heaviside}
		 \task Coulomb
		 \end{tasks} 
	 \item \emph{(\protect\tikz[baseline=-0.5ex]{\protect\draw[blue, very thick] (0,0) -- (0.5,0);})} The first indications of the study of electrostatics date back to the era of:
	 \begin{tasks}(4)
		 \task Phoenicians
		 \task \textbf{Greeks}
		 \task Egyptians
		 \task Romans
		 \end{tasks} 
	 \item \emph{(\protect\tikz[baseline=-0.5ex]{\protect\draw[blue, very thick] (0,0) -- (0.5,0);})} Who said the famous phrase “My battery is low and it’s getting dark”?
	 \begin{tasks}(4)
		 \task Earth Rover
		 \task \textbf{Opportunity Mars Rover}
		 \task Mars Rover
		 \task Discovery
		 \end{tasks} 
	 \item \emph{(\protect\tikz[baseline=-0.5ex]{\protect\draw[blue, very thick] (0,0) -- (0.5,0);})} Which scientist was rejected by his friend’s sister and was on the verge of suicide?
	 \begin{tasks}(4)
		 \task George Green
		 \task \textbf{William Rowan Hamilton}
		 \task Lewis Hamilton
		 \task Charles Leclerc
		 \end{tasks} 
	 \item \emph{(\protect\tikz[baseline=-0.5ex]{\protect\draw[blue, very thick] (0,0) -- (0.5,0);})} Who is considered the “prince of mathematics”?
	 \begin{tasks}(4)
		 \task \textbf{Carl Friedrich Gauss}
		 \task Terence Tao
		 \task Leonhard Euler
		 \task Antonio Alex Amor
		 \end{tasks} 
	 \item \emph{(\protect\tikz[baseline=-0.5ex]{\protect\draw[blue, very thick] (0,0) -- (0.5,0);})} What type of antennas resemble what Saiki Kusuo, protagonist of the comedy anime Saiki, has on his head?:
	 \begin{tasks}(4)
		 \task Aperture-type
		 \task \textbf{Dipoles}
		 \task Monopoles
		 \task Has nothing on his head
		 \end{tasks} 
	 \item \emph{(\protect\tikz[baseline=-0.5ex]{\protect\draw[blue, very thick] (0,0) -- (0.5,0);})} In the manga “Sousou no Frieren,” winner of the 14th Manga Taisho, which antennas are used for communication?
	 \begin{tasks}(4)
		 \task Dipoles
		 \task Yagi-Uda
		 \task Cassegrain
		 \task \textbf{Does not use antennas. Uses magic}
		 \end{tasks} 
	 \item \emph{(\protect\tikz[baseline=-0.5ex]{\protect\draw[blue, very thick] (0,0) -- (0.5,0);})} Who was the author of the idea of placing the golden records (the “Sounds of Earth”) showing humanity’s existence on the Voyager spacecraft?:
	 \begin{tasks}(4)
		 \task Einstein
		 \task Terence Tao
		 \task COS Sorzano
		 \task \textbf{Carl Sagan}
		 \end{tasks} 
	 \item \emph{(\protect\tikz[baseline=-0.5ex]{\protect\draw[blue, very thick] (0,0) -- (0.5,0);})} What was Sir Isaac Newton afraid of?:
	 \begin{tasks}(4)
		 \task \textbf{The dark}
		 \task His mother
		 \task His mother-in-law
		 \task Mathematics
		 \end{tasks} 
	 \item \emph{(\protect\tikz[baseline=-0.5ex]{\protect\draw[blue, very thick] (0,0) -- (0.5,0);})} Who primarily developed Yagi-Uda antenna? (Performed the practical work and experiments that established the antenna's conception): 
	 \begin{tasks}(4)
		 \task Yagi-Uda (1 person)
		 \task Hidetsugu Yagi
		 \task Uda-Yagi (1 person)
		 \task \textbf{Shintaro Uda}
		 \end{tasks} 
	 \item \emph{(\protect\tikz[baseline=-0.5ex]{\protect\draw[blue, very thick] (0,0) -- (0.5,0);})} What are the names of the two NASA satellites that chase each other to detect anomalies in Earth’s gravitational field?:
	 \begin{tasks}(4)
		 \task Itchy and Scratchy
		 \task Sylvester and Tweety
		 \task Wile E. Coyote and the Road Runner
		 \task \textbf{Tom and Jerry}
		 \end{tasks} 
	 \item \emph{(\protect\tikz[baseline=-0.5ex]{\protect\draw[blue, very thick] (0,0) -- (0.5,0);})} In the 12th century, magnets were also called:
	 \begin{tasks}(4)
		 \task Magnetar
		 \task Anuzkonio
		 \task Neodymium
		 \task \textbf{Lodestones}
		 \end{tasks} 
	 \item \emph{(\protect\tikz[baseline=-0.5ex]{\protect\draw[blue, very thick] (0,0) -- (0.5,0);})} Which portrait did Einstein NOT have in his house?:
	 \begin{tasks}(4)
		 \task Faraday's
		 \task \textbf{Lorentz's}
		 \task Maxwell's
		 \task Newton's
		 \end{tasks} 
	 \item \emph{(\protect\tikz[baseline=-0.5ex]{\protect\draw[blue, very thick] (0,0) -- (0.5,0);})} Which scientist was born later?
	 \begin{tasks}(4)
		 \task \textbf{Einstein}
		 \task Newton
		 \task Laplace
		 \task Marconi
		 \end{tasks} 
	 \item \emph{(\protect\tikz[baseline=-0.5ex]{\protect\draw[blue, very thick] (0,0) -- (0.5,0);})} The inventor of the monopole antenna was called:
	 \begin{tasks}(4)
		 \task \textbf{Guglielmo Marconi}
		 \task Marco Marconi
		 \task Emilio Marconi
		 \task Marconi Marconi
		 \end{tasks} 
	 \item \emph{(\protect\tikz[baseline=-0.5ex]{\protect\draw[blue, very thick] (0,0) -- (0.5,0);})} The alternating current motor was developed by:
	 \begin{tasks}(4)
		 \task Albert Einstein
		 \task Constantine Balanis
		 \task \textbf{Nikola Tesla}
		 \task Carl Friedrich Gauss
		 \end{tasks} 
	 \item \emph{(\protect\tikz[baseline=-0.5ex]{\protect\draw[blue, very thick] (0,0) -- (0.5,0);})} One of the main inventors of the telegraph has an important physical unit associated with his name:
	 \begin{tasks}(4)
		 \task Farad
		 \task \textbf{Henry}
		 \task Second
		 \task Ampere
		 \end{tasks} 
	 \item \emph{(\protect\tikz[baseline=-0.5ex]{\protect\draw[blue, very thick] (0,0) -- (0.5,0);})} The transmission line that connects the antenna to your television is typically:
	 \begin{tasks}(4)
		 \task Optical fiber
		 \task Copper pair
		 \task \textbf{Coaxial}
		 \task Microstrip
		 \end{tasks} 
	 \item \emph{(\protect\tikz[baseline=-0.5ex]{\protect\draw[blue, very thick] (0,0) -- (0.5,0);})} What type of radiation do antennas produce?:
	 \begin{tasks}(4)
		 \task Gamma radiation
		 \task Alpha radiation
		 \task Ionizing radiation
		 \task \textbf{Non-ionizing radiation}
		 \end{tasks} 
	 \item \emph{(\protect\tikz[baseline=-0.5ex]{\protect\draw[blue, very thick] (0,0) -- (0.5,0);})} Typical sector antennas for telephony consist of:
	 \begin{tasks}(4)
		 \task 1 sector
		 \task 2 sectors
		 \task \textbf{3 sectors}
		 \task 4 sectors
		 \end{tasks} 
	 \item \emph{(\protect\tikz[baseline=-0.5ex]{\protect\draw[blue, very thick] (0,0) -- (0.5,0);})} A solenoid generates uniform magnetic fields and has great applications in biomedicine in:
	 \begin{tasks}(4)
		 \task X-rays
		 \task Stethoscopes
		 \task \textbf{Magnetic resonance imaging (MRI)}
		 \task CT Scan
		 \end{tasks} 
	 \item \emph{(\protect\tikz[baseline=-0.5ex]{\protect\draw[blue, very thick] (0,0) -- (0.5,0);})} Router WiFi antennas typically operate at:
	 \begin{tasks}(4)
		 \task 2 GHz
		 \task \textbf{2.4 GHz}
		 \task 4.2 GHz
		 \task 4 GHz
		 \end{tasks} 
	 \item \emph{(\protect\tikz[baseline=-0.5ex]{\protect\draw[blue, very thick] (0,0) -- (0.5,0);})} The phrase “A scientific era ended and another began with James C. Maxwell” is attributed to:
	 \begin{tasks}(4)
		 \task Newton
		 \task \textbf{Einstein}
		 \task Ampère
		 \task Faraday
		 \end{tasks} 
	 \item \emph{(\protect\tikz[baseline=-0.5ex]{\protect\draw[blue, very thick] (0,0) -- (0.5,0);})} How many were Maxwell’s original equations as described by Maxwell himself?:
	 \begin{tasks}(4)
		 \task 4
		 \task 6
		 \task 10
		 \task \textbf{20}
		 \end{tasks} 
	 \item \emph{(\protect\tikz[baseline=-0.5ex]{\protect\draw[blue, very thick] (0,0) -- (0.5,0);})} Which scientist was born earlier?
	 \begin{tasks}(4)
		 \task Einstein
		 \task Lorentz
		 \task \textbf{Stokes}
		 \task Hertz
		 \end{tasks} 
	 \item \emph{(\protect\tikz[baseline=-0.5ex]{\protect\draw[blue, very thick] (0,0) -- (0.5,0);})} Which scientist associated with electromagnetism made important contributions to chemistry?:
	 \begin{tasks}(4)
		 \task \textbf{Faraday}
		 \task Heaviside
		 \task Balanis
		 \task Einstein
		 \end{tasks} 
	 \item \emph{(\protect\tikz[baseline=-0.5ex]{\protect\draw[blue, very thick] (0,0) -- (0.5,0);})} Who was one of the first people to demonstrate that light behaves as a wave?:
	 \begin{tasks}(4)
		 \task Einstein
		 \task Ørsted
		 \task \textbf{Maxwell}
		 \task Oppenheimer
		 \end{tasks} 
	 \item \emph{(\protect\tikz[baseline=-0.5ex]{\protect\draw[blue, very thick] (0,0) -- (0.5,0);})} Which metal has the greatest conductivity?
	 \begin{tasks}(4)
		 \task Iron
		 \task Copper
		 \task \textbf{Silver}
		 \task Gold
		 \end{tasks} 
	 \item \emph{(\protect\tikz[baseline=-0.5ex]{\protect\draw[blue, very thick] (0,0) -- (0.5,0);})} In the following anime, a helical antenna with primitive elements is designed:
	 \begin{tasks}(4)
		 \task To Aru Kagaku no Railgun
		 \task Mr. Robot
		 \task Naruto
		 \task \textbf{Dr. Stone}
		 \end{tasks} 
	 \item \emph{(\protect\tikz[baseline=-0.5ex]{\protect\draw[blue, very thick] (0,0) -- (0.5,0);})} Which antenna name is false?
	 \begin{tasks}(4)
		 \task Trombone
		 \task Rhombic antenna
		 \task V-antenna
		 \task \textbf{Wagner}
		 \end{tasks} 
	 \item \emph{(\protect\tikz[baseline=-0.5ex]{\protect\draw[blue, very thick] (0,0) -- (0.5,0);})} Professor C. Balanis was born in:
	 \begin{tasks}(4)
		 \task United States
		 \task Italy
		 \task \textbf{Greece}
		 \task France
		 \end{tasks} 
	 \item \emph{(\protect\tikz[baseline=-0.5ex]{\protect\draw[blue, very thick] (0,0) -- (0.5,0);})} The Yagi-Uda antenna:
	 \begin{tasks}(4)
		 \task \textbf{It is an evolution of the dipole}
		 \task It is an aperture antenna
		 \task It is a frequency-independent antenna
		 \task It is an evolution of microstrip
		 \end{tasks} 
	 \item \emph{(\protect\tikz[baseline=-0.5ex]{\protect\draw[blue, very thick] (0,0) -- (0.5,0);})} Which scientist was president of the Royal Society?
	 \begin{tasks}(4)
		 \task Tesla
		 \task \textbf{Stokes}
		 \task Gauss
		 \task Ampère
		 \end{tasks} 
	 \item \emph{(\protect\tikz[baseline=-0.5ex]{\protect\draw[blue, very thick] (0,0) -- (0.5,0);})} To watch satellite TV in Spain, the antenna typically used is:
	 \begin{tasks}(4)
		 \task Short dipole
		 \task Yagi-Uda antenna
		 \task \textbf{Parabolic antenna}
		 \task Fractal antenna
		 \end{tasks} 
	 \item \emph{(\protect\tikz[baseline=-0.5ex]{\protect\draw[blue, very thick] (0,0) -- (0.5,0);})} Hans Christian Oersted, one of the modern-era scientists associated with electromagnetism, was:
	 \begin{tasks}(4)
		 \task \textbf{Danish}
		 \task French
		 \task English
		 \task Andorran
		 \end{tasks} 
	 \item \emph{(\protect\tikz[baseline=-0.5ex]{\protect\draw[blue, very thick] (0,0) -- (0.5,0);})} The industry prefers to send antennas into outer space that only contain metal (no dielectric) because:
	 \begin{tasks}(4)
		 \task They are cheaper
		 \task They are prettier
		 \task \textbf{Metals and dielectrics expand non-uniformly with temperature}
		 \task False
		 \end{tasks} 
	 \item \emph{(\protect\tikz[baseline=-0.5ex]{\protect\draw[blue, very thick] (0,0) -- (0.5,0);})} Retarded potentials are also known as:
	 \begin{tasks}(4)
		 \task Planck potentials
		 \task Ampère–Faraday potentials
		 \task \textbf{Liénard–Wiechert potentials}
		 \task Maxwell–Planck potentials
		 \end{tasks} 
	 \item \emph{(\protect\tikz[baseline=-0.5ex]{\protect\draw[blue, very thick] (0,0) -- (0.5,0);})} Which antenna name is false?
	 \begin{tasks}(4)
		 \task Hertzian dipole
		 \task Franklin antenna
		 \task Yagi antenna
		 \task \textbf{Kafka antenna}
		 \end{tasks} 
	 \item \emph{(\protect\tikz[baseline=-0.5ex]{\protect\draw[blue, very thick] (0,0) -- (0.5,0);})} Which scientist first observed the deflection of a compass by the passage of a current?
	 \begin{tasks}(4)
		 \task Faraday
		 \task Ampère
		 \task \textbf{Oersted}
		 \task Gauss
		 \end{tasks} 
	 \item \emph{(\protect\tikz[baseline=-0.5ex]{\protect\draw[blue, very thick] (0,0) -- (0.5,0);})} Radio waves that pass through the ionosphere rotate their polarization plane. This effect is known as:
	 \begin{tasks}(4)
		 \task \textbf{Faraday effect}
		 \task Alex effect
		 \task Ampère effect
		 \task Hertz effect
		 \end{tasks} 
	 \item \emph{(\protect\tikz[baseline=-0.5ex]{\protect\draw[blue, very thick] (0,0) -- (0.5,0);})} The fish-eye lens (antenna) were described and designed by:
	 \begin{tasks}(4)
		 \task Antonio Vivaldi
		 \task \textbf{Maxwell}
		 \task Hertz
		 \task Marconi
		 \end{tasks} 
	 \item \emph{(\protect\tikz[baseline=-0.5ex]{\protect\draw[blue, very thick] (0,0) -- (0.5,0);})} Who developed the first antenna (dipole)?:
	 \begin{tasks}(4)
		 \task Faraday
		 \task \textbf{Hertz}
		 \task Lorentz
		 \task Marconi
		 \end{tasks} 
	 \item \emph{(\protect\tikz[baseline=-0.5ex]{\protect\draw[blue, very thick] (0,0) -- (0.5,0);})} Which antenna has the ability to precisely focus rays at one of its foci?:
	 \begin{tasks}(4)
		 \task Spherical reflector
		 \task Plane reflector
		 \task \textbf{Parabolic reflector}
		 \task Hyperbolic reflector
		 \end{tasks} 
	 \item \emph{(\protect\tikz[baseline=-0.5ex]{\protect\draw[blue, very thick] (0,0) -- (0.5,0);})} The main antenna mounted on old car roofs is:
	 \begin{tasks}(4)
		 \task \textbf{Quarter-wave monopole}
		 \task Half-wave dipole
		 \task Yagi $\lambda$/2
		 \task Quadrupole
		 \end{tasks} 
	 \end{enumerate}

\end{document}